\documentclass[fleqn,usenatbib]{mnras}

\usepackage[T1]{fontenc}
\usepackage{ae,aecompl}
\usepackage{xcolor}

\usepackage{graphicx}	
\usepackage{amsmath}	
\usepackage{amssymb}	

\usepackage{eso-pic}

\AddToShipoutPictureBG*{%
  \AtPageUpperLeft{%
    \hspace{0.75\paperwidth}%
    \raisebox{-3.5\baselineskip}{%
      \makebox[0pt][l]{\textnormal{DES-2020-0568}}
}}}%

\AddToShipoutPictureBG*{%
  \AtPageUpperLeft{%
    \hspace{0.75\paperwidth}%
    \raisebox{-4.5\baselineskip}{%
      \makebox[0pt][l]{\textnormal{FERMILAB-PUB-20-533-AE}}
}}}%

\usepackage{newtxtext,newtxmath}

\def\ptilde{\mathrm{\tilde{p}}}
\def\petilde{\mathrm{\tilde{p}_e}}

\title[DES morphological catalog]{Pushing automated morphological classifications to their limits with the Dark Energy Survey}

\author[J. Vega-Ferrero et al.]{
\parbox{\textwidth}{
\Large
J.~Vega-Ferrero,$^{1,2}$\thanks{E-mail: vegaf@sas.upenn.edu} 
H.~Dom\'{\i}nguez S\'anchez,$^{1,3}$
M.~Bernardi,$^{1}$
M.~Huertas-Company,$^{4,5,6,7,8}$
R.~Morgan,$^{9}$
B.~Margalef,$^{1}$
M.~Aguena,$^{10,11}$
S.~Allam,$^{12}$
J.~Annis,$^{12}$
S.~Avila,$^{13}$
D.~Bacon,$^{14}$
E.~Bertin,$^{15,16}$
D.~Brooks,$^{17}$
A.~Carnero~Rosell,$^{7,18}$
M.~Carrasco~Kind,$^{19,20}$
J.~Carretero,$^{21}$
A.~Choi,$^{22}$
C.~Conselice,$^{23,24}$
M.~Costanzi,$^{25,26}$
L.~N.~da Costa,$^{11,27}$
M.~E.~S.~Pereira,$^{28}$
J.~De~Vicente,$^{29}$
S.~Desai,$^{30}$
I.~Ferrero,$^{31}$
P.~Fosalba,$^{32,3}$
J.~Frieman,$^{12,33}$
J.~Garc\'ia-Bellido,$^{13}$
D.~Gruen,$^{34,35,36}$
R.~A.~Gruendl,$^{19,20}$
J.~Gschwend,$^{11,27}$
G.~Gutierrez,$^{12}$
W.~G.~Hartley,$^{37,17,38}$
S.~R.~Hinton,$^{39}$
D.~L.~Hollowood,$^{40}$
K.~Honscheid,$^{22,41}$
B.~Hoyle,$^{42,43,44}$
M.~Jarvis,$^{1}$
A.~G.~Kim,$^{45}$
K.~Kuehn,$^{46,47}$
N.~Kuropatkin,$^{12}$
M.~Lima,$^{10,11}$
M.~A.~G.~Maia,$^{11,27}$
F.~Menanteau,$^{19,20}$
R.~Miquel,$^{48,21}$
R.~L.~C.~Ogando,$^{11,27}$
A.~Palmese,$^{12,33}$
F.~Paz-Chinch\'{o}n,$^{49,20}$
A.~A.~Plazas,$^{50}$
A.~K.~Romer,$^{51}$
E.~Sanchez,$^{29}$
V.~Scarpine,$^{12}$
M.~Schubnell,$^{28}$
S.~Serrano,$^{32,3}$
I.~Sevilla-Noarbe,$^{29}$
M.~Smith,$^{52}$
E.~Suchyta,$^{53}$
M.~E.~C.~Swanson,$^{20}$
G.~Tarle,$^{28}$
F.~Tarsitano,$^{38}$
C.~To,$^{34,35,36}$
D.~L.~Tucker,$^{12}$
T.~N.~Varga,$^{43,44}$
and R.D.~Wilkinson$^{51}$
}
  \vspace{0.4cm}\\~\\
\parbox{\textwidth}{\centering \textsc{\Large  } \\ \centering \textit{Author affiliations are listed at the end of this paper} }
\vspace{0.4cm}
}

\date{Accepted 2021 February 22. Received 2021 February 5; in original form 2020 October 21}

\pubyear{2021}

\begin{document}
\label{firstpage}
\pagerange{\pageref{firstpage}--\pageref{lastpage}}
\maketitle

\newcommand\ls{\ensuremath{\hbox{\rlap{\hbox{\lower4pt\hbox{$\sim$}}}\hbox{$<$}}}}
\newcommand\gs{\ensuremath{\hbox{\rlap{\hbox{\lower4pt\hbox{$\sim$}}}\hbox{$>$}}}}

\begin{abstract}

We present morphological classifications of $\sim$27 million galaxies from the Dark Energy Survey (DES) Data Release 1 (DR1) using a supervised deep learning algorithm. The classification scheme separates: 
  (a) early-type galaxies (ETGs) from late-types (LTGs); and 
  (b) face-on galaxies from edge-on.  
Our Convolutional Neural Networks (CNNs) are trained on a small subset of DES objects with previously known classifications.  These typically have $\mathrm{m}_r \lesssim 17.7~\mathrm{mag}$; we model fainter objects to $\mathrm{m}_r < 21.5$ mag by simulating what the brighter objects with well determined classifications would look like if they were at higher redshifts.  The CNNs reach 97\% accuracy to $\mathrm{m}_r<21.5$  on their training sets, suggesting that they are able to recover features more accurately than the human eye.  We then used the trained CNNs to classify the vast majority of the other DES images.  The final catalog comprises five independent CNN predictions for each classification scheme, helping to determine if the CNN predictions are robust or not. We obtain secure  classifications for $\sim$ 87\% and  73\% of the catalog for the  ETG vs. LTG and edge-on vs. face-on models, respectively. Combining the two classifications (a) and (b) helps to increase the purity of the ETG sample and to identify edge-on lenticular galaxies (as ETGs with high ellipticity). Where a comparison is possible, our classifications correlate very well with S\'ersic index (\textit{n}), ellipticity ($\epsilon$) and spectral type, even for the fainter galaxies. This is the largest  multi-band  catalog of automated galaxy morphologies to date.

\end{abstract}

\begin{keywords}
methods: observational, catalogues, galaxies: structure
\end{keywords}


\section{Introduction}\label{sec:intro}

Morphology is a fundamental property of a galaxy. It is intimately related to galaxy mass, star formation rate (SFR), stellar kinematics and environment \citep[e.g.][]{ Pozzetti2010, Wuyts2011, Huertas-Company2016}. Galaxy structure changes across cosmic time and this mutation is intertwined with formation channels and evolutionary paths. Whether galaxy morphology determines the fate of a galaxy or, conversely, morphological transformations are driven by their stellar population content is still a matter of debate \citep[e.g.][]{Lilly2016}. Distinguishing between the two options is one of the main challenges in understanding galaxy formation and evolution.

Having large samples of galaxies with morphological classifications is crucial for studying the relation between shapes and star formation histories or mass assembly. Traditionally, morphological classification was done by visual inspection. This method has the great inconvenience of being very expensive in terms of time (limiting the available samples to a few thousands -- e.g., \citealt{Nair2010}), but it is also affected by the subjectivity of the classifier. An alternative is people-powered research like the Galaxy Zoo project \citep[e.g.][]{Lintott2008, Willett2013, Walmsley2020}, where volunteers are asked to classify galaxies. The large number of classifiers significantly reduces the task time and allows for a statistical analysis of the answers. However, these methods still present important biases (see figure 24 in \citealt{Fischer2019}).  More importantly, they will be unable to keep up with the enormous amount of data (millions of galaxy images) that the next generation of surveys such as the Vera Rubin Observatory Legacy Survey of Space and Time or Euclid will deliver: about a hundred years would be needed to classify all data from the Euclid mission with a Galaxy Zoo-like approach. Therefore, applying automated classification methods to such large surveys is mandatory.

An alternative common approach is the quantitative estimation of galaxy morphology. In this methodology, the galaxy light is described in terms of structural quantities (e.g., magnitude, size, ellipticity, asymmetry, concentration, etc.). For DES, such measurements are available for $\sim$50 million galaxies up to $\mathrm{m}_i$ $ <$ 23 mag \citep[][which also provides a comprehensive overview]{Tarsitano2018}. This technique can rely either on the parametrization of galaxy light profile (e.g., S\'ersic function) or on the analysis of the distribution of galaxy light without accounting for the PSF. Therefore it requires specific calibrations.

Recent studies in machine learning and deep learning techniques, in particular, present an attractive way forward. The use of convolutional neural networks (CNN) for the analysis of galaxy images has proven to be extremely successful for classifying galaxy images \citep[e.g.][-- and many others]{Dieleman2015, Aniyan2017, Tuccillo2018, Huertas-Company2018, Dominguez2018, Metcalf2019, Pasquet2019, Ntampaka2019, Hausen2020, Ghosh2020}. CNNs have overtaken other supervised automated methods such as Logistic Regression, Support Vector Machines, random forest, decision trees, etc., in terms of both accuracy and computation time (see \citealt{Cheng2020} for a detailed comparison), specially for image-based (or array-based) data. However, supervised algorithms rely on large samples of pre-labelled objects on which to train. Complex classification, such as the separation between ETGs and LTGs, requires deep CNNs with a large number of free parameters. 
These training samples should come from the same data domain (e.g., instrument, depth) as the sample to be classified. This ideal is particularly challenging for new surveys where the overlap with available morphological catalogues may be limited. Transfer learning between surveys is an alternative that helps reduce the required training sample size by almost one order of magnitude (see \citealt{DS2019}), but  a set of labelled objects with a similar distribution to the main target sample is still needed. 

In this context, we aim to provide morphological classifications for galaxies from the Dark Energy Survey public data release DR1  \citep{DES2018}. Although the scope of  DES  is to probe the nature of dark energy, the survey has observed the sky over 6 years mapping $\sim$~300 million galaxies in the first three years as a by-product of DR1 -- which will become $\sim$ 600 million galaxies for DR2 \citep{DES6}. This dataset is one of the largest and deepest among modern galaxy surveys, reaching up to 24~mag in the \textit{r}-band. Although there are enough DES galaxies in common with previous morphological catalogues (in particular with \citealt{Dominguez2018}) to properly train CNNs, they are limited to bright magnitudes ($\mathrm{m}_r < 17.7~\mathrm{mag}$). 

To reach the fainter magnitudes that are necessary to probe the redshift evolution of morphological transformations, in section  \ref{sec:data} we use DES galaxies with well-known classifications and simulate what they would look like if they were at higher redshifts. This dramatically reduces the quality of the redshifted images (while keeping track of the original \textit{true} labels). We then check if the CNNs are able to recover features hidden to the human eye.  In section \ref{sect:models} we use the original and simulated samples to train our CNNs to classify images as  ETGs or LTGs, and edge-on or face-on.  We then compare our CNN classifications with the corresponding \textit{true} labels of a sub-sample that was reserved for testing, as well as with the properties of faint DES galaxies from other available catalogues (see section \ref{sec:results}). 

This is the largest catalogue of galaxy morphological classification to date (as detailed in section \ref{sec:des_morph}), along with the independent catalog produced by the companion DES paper presented in Cheng et al., where CNNs are used to classify DES galaxies into elliptical or spirals on the basis of the $i$-band image. The multiband morphological catalog presented in this work provides reliable ETG/LTG classifications for 27 million galaxies from the DES survey with $\mathrm{m}_r < 21.5~\mathrm{mag}$. Our catalog also includes an edge-on classification, which  can be useful for other science analyses (e.g., probing self interacting dark matter, \citealt{Secco2018}; estimating dust attenuation, \citealt{Li2020,Masters2010}; or stuying diffuse ionized gas, \citealt{Levy2019}; among others).


\section{Data Sets}
\label{sec:data}

In this section, we describe the dataset used for training and testing, as well as the final sample to which we apply our models in the construction of our catalogue.


\subsection{Dark Energy Survey science DR1}
\label{sec:desy3}

The main objective of this work is to provide morphological classifications for a large set of galaxies in the public release of the first three years of DES data\footnote{DES database is publicly aaccesible at \url{https://des.ncsa.illinois.edu/desaccess/}} (DES DR1, \citealt{DES2018}). The DES DR1 covers the full DES footprint (about 5,000 square degrees) and includes roughly 40,000 exposures taken with the Dark Energy Camera \citep{Flaugher2015}. The coadd images, taken in \textit{griz}-bands, are available along with catalogs of roughly 300 million galaxies, reaching a median co-added catalog depth of $g$ = 24.33, $r$ = 24.08, $i$ = 23.44 {mag} at signal-to-noise ratio S/N = 10, with a pixel size of 0.263\arcsec (see \citealt{DES2018} for technical details).


We selected a high quality galaxy sample based on a classification that separates PSF-like objects (such as stars and QSOs) and extended objects (i.e., galaxies). The classifier is denoted as EXTENDED\_CLASS\_COADD in the DES database and it is derived using the \textit{spread\_model} quantity from Sextractor photometry \citep[see][for more details]{Abbott2018}; its value should be greater than 1 in order to select medium and high confidence galaxies. We also excluded regions of the sky 
with missing data or bright stars in any of the observed bands  by employing the masks described in  DES database, since that could affect our model predictions.  We used photometric data in the \textit{gri}-bands and selected galaxies brighter than $\mathrm{m}_r = 21.5~\mathrm{mag}$, where $\mathrm{m}_r$ denotes the magnitude in an elliptical aperture shaped by Kron radius in the \textit{r}-band (MAG\_AUTO\_R in the DES database), and with a half-light radius in the \textit{r}-band (FLUX\_RADIUS\_R in the DES database; denoted  as $\mathrm{r}_r$ throughout the paper) larger than 2.8 pixels (or 0.74 \arcsec; see section~\ref{sec:network}). See table~\ref{tab:catalog} for more details. This selection produces a final catalog of 26,971,945 (i.e., nearly 27 million) galaxies. We provide morphological classifications for these galaxies and describe our catalog in Section \ref{sec:des_morph}.


\subsection{SDSS morphological catalogue}
\label{sec:ds18}

To derive morphological classifications of galaxies within the DES footprint, we have used the morphological catalog published by  \citet[][DS18 hereafter]{Dominguez2018}, which partially overlaps with the DES DR1 (see section \ref{sec:training} for details).  The DS18 is a publicly available catalogue that provides morphologies for $\sim$670,000 galaxies in the Sloan Digital Sky Survey (SDSS) with $\mathrm{m}_r < 17.7~\mathrm{mag}$.  These were obtained by combining accurate existing visual classification catalogues (\citealt{Nair2010}) and Galaxy Zoo 2 (\citealt{Willett2013}; hereafter GZOO) with deep learning (DL) algorithms using CNNs. The DS18 catalogue provides several GZOO-like classifications, as well as a T-Type (i.e. the numerical Hubble stage, \citealt{deVaucouleurs1959}) and a separation between elliptical and S0 galaxies.  Although these classifications are automated (i.e., derived without any visual inspection), they are good enough (accuracy > 97\%) to provide reliable labels for our training sample. 


\subsection{Simulating DES galaxies at higher redshifts}
\label{sec:sims}

Although the number of DES galaxies that are also in the DS18 catalogue is large ($\sim$20,000), it is unlikely that a CNN trained on a galaxy sample brighter than $\mathrm{m}_r < 18$ mag would accurately classify the vast majority of considerably fainter galaxies in DES. One way to remedy this issue is to visually classify a sample of fainter galaxies. That would have been a tedious task, subject to biases, since different classifiers (i.e., observers) would almost certainly assign different classifications, (as can be seen in the GZOO catalog). In addition, some of the features that distinguish ETGs from LTGs are not so evident for faint distant galaxies (e.g. spiral arms, bar; see figure \ref{fig:high-z_examples}), which would complicate the classification. An alternative to visual inspection is to simulate what actual DES galaxies with a well-known classification would look like if they were at higher redshift. We simulate the effects of observing a galaxy at a higher $z$ given its original DES cutout at $z_0$. To do so we use \texttt{GALSIM} \footnote{https://github.com/GalSim-developers/GalSim} \citep{Rowe2015} and assume a $\Lambda$CDM cosmology with $\Omega_M=0.3$,  $\Omega_{\Lambda}=0.7$ and $h=0.7$. We perform the following steps:

\begin{figure*}
\centering
 \includegraphics[width=2\columnwidth]{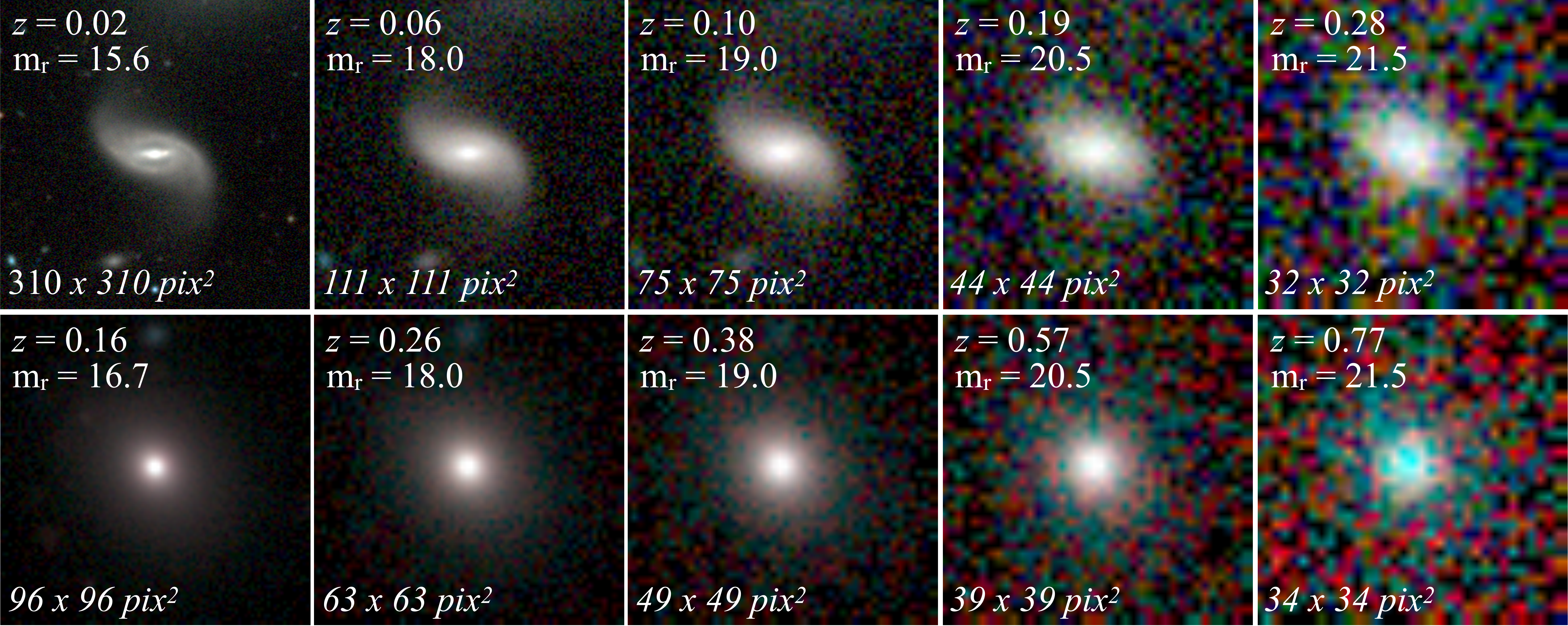}
 \caption{Cutouts of a LTG (upper panels) with T-Type $=5.0$ from DS18 observed at $z_0 = 0.02$ with $\mathrm{m}_r = 15.6~\mathrm{mag}$ and for an ETG (lower panels) with T-Type $=-2.4$ observed at $z_0 = 0.16$ and $\mathrm{m}_r = 16.7~\mathrm{mag}$. Cutouts from left to right show the original galaxies redshifted to an apparent magnitude of $\mathrm{m}_r = (18.0, 19.0, 20.5, 21.5)~\mathrm{mag}$, respectively. For each panel, the redshift ($z$) and the apparent magnitude ($\mathrm{m}_r$) are shown on the upper left corner, while the size of each image (in pixels) is indicated in the lower left corner. The features that distinguish ETG and LTG galaxies become less evident at fainter magnitudes.
\label{fig:high-z_examples}}
\end{figure*}

\begin{itemize}

    \item a de-convolution of the original image by the corresponding point spread function (PSF) in each band. As an approximation for the PSF\footnote{This approximation has no impact in the performance of the classifications on real DES images, as demonstrated in section~\ref{sec:results}.}, we assume a Moffat surface brightness profile ($\beta = 2$) with FWHM equal to the FWHM values for the DES PSF presented in \citet{Abbott2018}, $(1.12,0.96,0.88)$ arcsec for the \textit{gri}-bands, respectively;
    
    \item a change in angular size due to the cosmological dimming as follows:
    \begin{equation}
        \frac{\mathrm{s}(z)}{\mathrm{s}(z_\mathrm{0})} = \frac{\mathrm{D_L}(z_\mathrm{0})}{\mathrm{D_L}(z)} \frac{(1+z)^2}{(1+z_\mathrm{0})^2},
    \end{equation}
    where $\mathrm{s}$ denotes the size of the image, $\mathrm{D_L}$ corresponds to the luminosity distance and $z_\mathrm{0}$ and $z$ are the observed and the simulated redshift, respectively. Note that we keep constant the DES pixel size of $0.263\arcsec$ and, therefore, it is the size of the image in pixels that changes (it shrinks with increasing $z$);
    
    \item a change in the apparent magnitude. This change includes the cosmological dimming effect, the $k$-correction and the evolution of the intrinsic brightness of the galaxies. For the ETGs, we assume the evolution corresponds to that of a single stellar population model, while for the LTG class we assume a constant star formation rate (SFR, taking the value from SDSS spectroscopy).  In summary, we express the change in apparent magnitude as:
    
    \begin{equation}
        \mathrm{m}(z) - \mathrm{m}(z_\mathrm{0}) = \Delta\mathrm{m_{evo}} + ~5\mathrm{log} \left[ \frac{\mathrm{D_L}(z)}{\mathrm{D_L}(z_\mathrm{0})} \right],
    \end{equation}
    where $\mathrm{m}(z_\mathrm{0})$ and $\mathrm{m}(z)$ indicate the observed and redshifted apparent magnitude, respectively, and $\Delta\mathrm{m_{evo}}$ corresponds to the change in magnitude according to the $k$-correction and the evolutionary models;

    \item a convolution of the resulting image by the above mentioned PSF in each band;
    
    \item  the addition of Gaussian noise to the final image. In order to avoid contamination from the central galaxy, we estimate the noise from a set of cutouts with a larger field of view ($\sim 20$ times the half-light radius). We use a robust wavelet-based estimator of the Gaussian noise standard deviation (\texttt{estimate\_sigma} function available for the \texttt{scikit-image}\footnote{https://scikit-image.org/} package in \texttt{Python}).
    
\end{itemize}

We apply this procedure to each band independently and then we combine the three bands into an RGB image. We simulate each galaxy satisfying the following conditions: a) the final apparent magnitude (in the \textit{r}-band) below  $\mathrm{m}_r(z) < 22.5$; b) the final size of the image, $\mathrm{s}(z)$, larger than $32\times32$ pixels; c) and the final redshift  $z$ < 1.0. The first condition ensures that the CNN is learning from images of galaxies that are even fainter than the limiting brightness of our project ($\mathrm{m}_r < 21.5$) but are still bright enough to pass the DES detection threshold. The second condition avoids extreme interpolations when constructing the input matrix (see section \ref{sec:network}).  As mentioned above, the pixel size is kept constant, while the size of the image decreases with increasing $z$. This choice ensures that both the original and the simulated image of the galaxy have the same physical size.


 In figure~\ref{fig:high-z_examples}, we show the original cutout at $z_0$ along with four simulated cutouts for two galaxies (one LTG and one ETG). For the LTG galaxy (T-Type $=5.0$ from DS18), it can be clearly seen how the spiral arms and the bar are almost indistinguishable (by eye) when the galaxy is simulated at a $\mathrm{m}_r \geq 20.5~\mathrm{mag}$. The original size of the LTG image is $310\times310$ pixels, while its simulation at $\mathrm{m}_r = 21.5~\mathrm{mag}$ is only $32\times32$ pixels. For the ETG galaxy (T-Type $=-2.4$), the original size of the image is $96\times96$ pixels, while the size of its simulation at $\mathrm{m}_r = 21.5~\mathrm{mag}$ is $34\times34$ pixels.

 
\section{Deep Learning morphological classification model}
\label{sect:models}

\begin{figure*}
\centering
\includegraphics[width=0.9\textwidth]{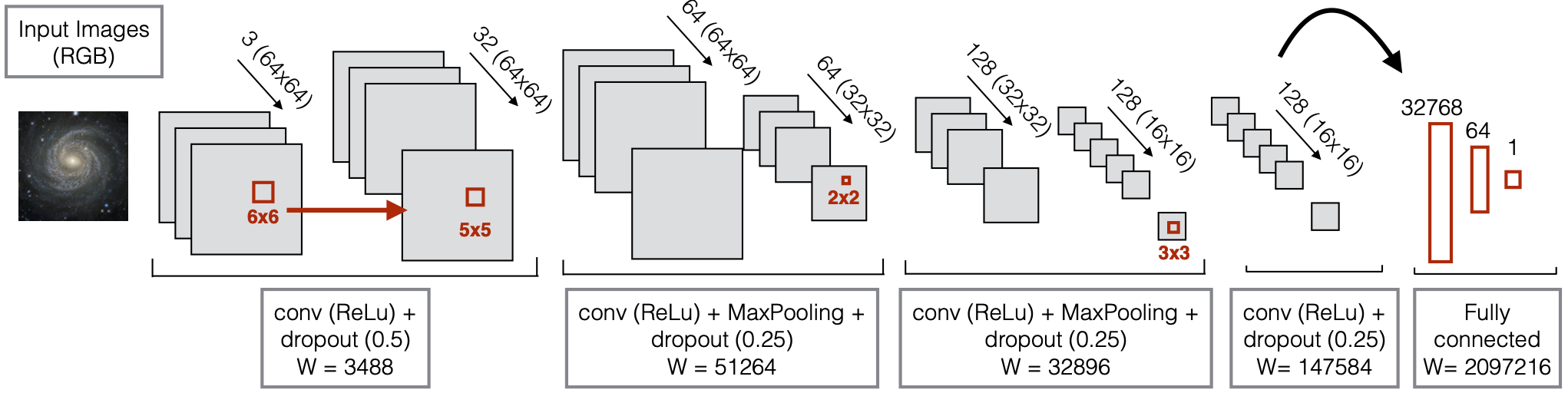}
\caption{Network architecture used for training the models, consisting of four convolutional layers with a linear activation function (denoted as \textit{ReLu}) and different kernel sizes represented by the red squares, followed by a fully connected layer. The numbers above each layer correspond to the size of the output convoluted images, while the number of weights at each level are indicated below and denoted by W.}
\label{fig:network} 
\end{figure*}

We apply DL algorithms using CNNs to morphologically classify galaxy images. DL is a methodology that automatically learns and extracts the most relevant features (or parameters) from raw data for a given classification problem through a set of non-linear transformations. The main advantage of this methodology  is that no pre-processing needs to be done: the input to the machine are the raw RGB cutouts for each galaxy (i.e., $gri$-bands, respectively). The main disadvantage is that, given the complexity of extracting and optimising the features and weights in each layer, a large number of already classified (or labelled) images needs to be provided to the machine. As explained in section \ref{sec:data}, we combine a previous morphological catalogue that overlaps with the DES dataset with simulations of the original DES images for training and testing how our model performs. 

Since we want to apply the morphological classification to the DES sample that covers a much fainter range of objects than the SDSS sample, we limit the morphological classification to a simpler scheme than that presented in DS18: we classify the DES galaxies according to two different schemes: a) early-type galaxies (ETGs) vs. late-type galaxies (LTGs); and b) face-on vs. edge-on galaxies.


\subsection{Network architecture}
\label{sec:network}

Given the success of previous studies that have used CNNs for galaxy classification \citep{Huertas2015, Dieleman2015, Dominguez2018}, we adopt a similar (but not identical) CNN configuration. Testing the performance of different network architectures is beyond the scope of this paper.  We use the same input images and CNN configuration for each classification task. We use the  \texttt{KERAS} library\footnote{https://keras.io/}, a high-level neural network application programming interface, written in  \texttt{Python}.

The input to the CNN are the RGB cutouts (i.e., $gri$-bands) downloaded from the DES DR1, with a varying size that is function of the half-light radius of the galaxy in the \textit{r}-band. The cutouts have a size of $\sim 11.4$ times the half-light radius centred on the target galaxy to guarantee that no galaxy light is missed. The algorithm reads the images that are re-sampled into (64, 64, 3) matrices, with each number representing the flux in a given pixel at a given band. Down-sampling the input matrix is necessary to reduce the computing time and to avoid over-fitting in the models, as commonly done in the literature \citep{Dieleman2015,Dominguez2018,Walmsley2020}. The flux values are normalised to the maximum value in each filter for each galaxy to eliminate colour information. For the smaller galaxies, the fixed pixel size can lead to cutout sizes that are below the $64\times64$ pixels for which the CNN has been designed. For these, the cutouts are up-sampled to 64$\times$64 matrices by interpolating between pixels. Since this could create some artifacts and affect the spatial resolution of the images, we require all cutouts to be at least $32\times32$ pixels in size. This condition leads to a minimum galaxy half-light radius of 2.8 ($32/11.4$) pixels (as mentioned in section~\ref{sec:desy3}). 

The network architecture, shown in figure \ref{fig:network}, is composed of four convolutional layers with a linear activation function (denoted as \textit{ReLu}) and squared filters of different sizes (6, 5, 2 and 3, respectively), followed by a fully connected layer. Dropout is performed after each convolutional layer to avoid over-fitting, and a 2$\times$2 max-pooling follows the second and third convolutional layers. The number of weights (i.e., free-parameters) in each layer -- before dropout -- are also indicated. \citep[See][ for a comprehensive review on DL concepts]{Goodfellow2016}.


We train the models in  binary classification mode for both classification schemes. For each, the output is a single value between 0 and 1, and can be interpreted as the probability $\mathrm{p}$ of being a positive example (labelled as $\mathrm{Y}=1$ in our input matrix) or as the probability $1-\mathrm{p}$ of being a negative example (labelled as $\mathrm{Y}=0$ in our input matrix). We use 50 training epochs, with a batch size of 30 and an \textit{adam} optimization (default \textit{learning rate} of 0.001). In the training process, we perform  \textit{data augmentation}, allowing the images to be zoomed in and out (0.75 to 1.3 times the original size), flipped and shifted  both vertically and horizontally (by 5\%). This ensures our model does not suffer from over-fitting since the input is not the same in every training epoch. \textit{Early stopping} is also switched on with a maximum of 10 epochs after convergence is reached. The best model, defined as the optimal one for the validation sample during the training, is then saved and applied to cutouts from the full DES DR1 galaxy catalog, generated in the same way as for the training sample.


\subsection{Training sample}
\label{sec:training}

\subsubsection{Primary training sample}

Our primary training sample is the result of cross-matching the sources in the DS18 and the DES DR1 catalogs presented in section~\ref{sec:data}. We identified sources in both catalogs as those with a separation in the sky of less than 2 arcsec, after removing multiple identifications. We remove those objects missing spectra (or having bad spectroscopy according to SDSS flags) and with relative differences in $z$ of more than 5\% between the photo-$z$ for DES Y3 Gold catalog (\citealt{deVicente2016} and Sevilla-Noarbe et al. in prep.) and spec-$z$ for SDSS data. Only 50 galaxies are excluded according to these criteria. The resulting catalog consists of 19,913 galaxies with good quality imaging and secure spectroscopic $z$.

\subsubsection{Simulated training sample}

The DS18 catalog (described in section~\ref{sec:ds18}) only reaches an observed magnitude of $\mathrm{m}_r < 17.7$ mag. Since we aim to push the limits of the morphological classification of galaxies to fainter magnitudes, we extend our primary training sample by simulating each galaxy at higher redshift $z$ and, consequently, making it look fainter and smaller.

\begin{figure}
\centering
 \includegraphics[width=\columnwidth]{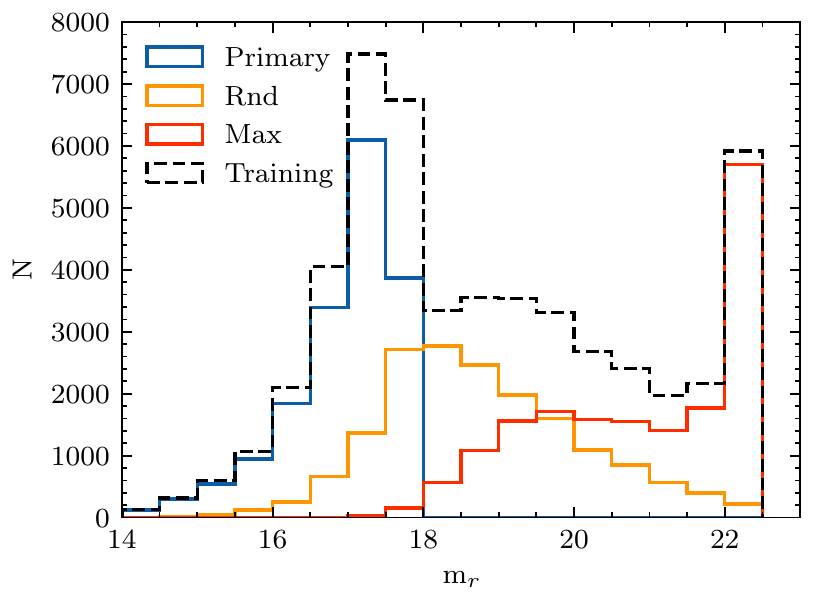}
 \includegraphics[width=0.975\columnwidth]{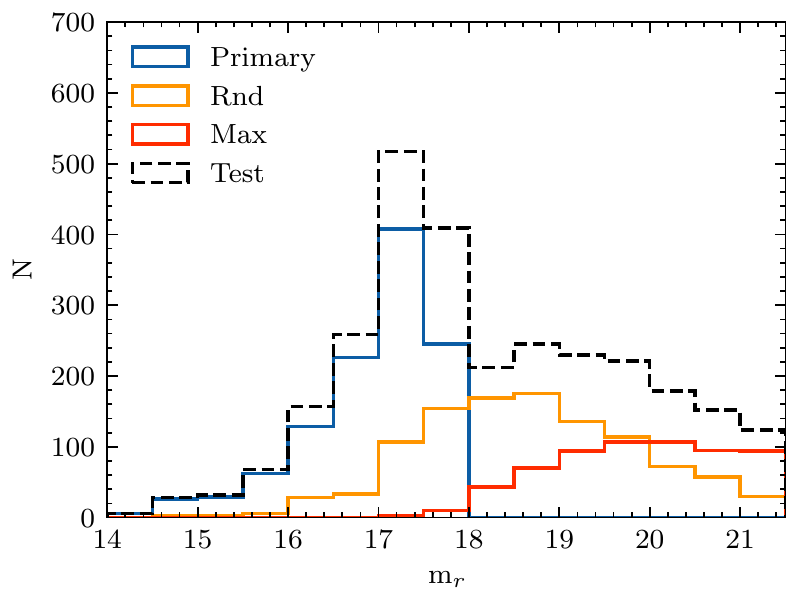}
 \caption{Top panel: Distribution of apparent magnitude in the \textit{r}-band ($\mathrm{m}_r$) for the primary training sample (solid blue); the two sets of simulations (solid orange and red show the \textit{rnd} and \textit{max} sets); and the training sample distribution (dashed black) used for the ETG vs. LTG classification scheme.  Bottom panel: Same as top but only for the test sample (see section~\ref{sec:ETG_vs_LTG}). Note that the CNN predictions are trained with a sample that extends to $\mathrm{m}_r < 22.5$, while the model is tested only to $\mathrm{m}_r < 21.5$ (the limit of the DES catalogue presented in this work).
\label{fig:appmag}}
\end{figure}

Following the pipeline described in section \ref{sec:sims}, we generate two sets of simulations: 
 a) one at a random $z$ (hereafter \textit{rnd}) chosen from a uniform distribution between the observed $z_{\mathrm{0}}$ and the maximum $z_{max}$ to which the galaxy can be redshifted according to the criteria mentioned in section \ref{sec:sims} (i.e.,  brighter than $\mathrm{m}_r(z) < 22.5$; cutout larger than $32\times32$ pixels and $z$ < 1); 
 b) a second one at the maximum $z$ (hereafter \textit{max}) under the three given conditions above. 
By combining the primary training sample and the two sets of simulations we obtain a more uniform distribution of the apparent magnitude as can be seen in figure~\ref{fig:appmag} for the \textit{r}-band.  Note that, the primary training set is limited in apparent magnitude to $\mathrm{m}_r < 17.7~\mathrm{mag}$, while the two set of simulations extend our training sample to $\mathrm{m}_r < 22.5~\mathrm{mag}$. As there are indications that close to the limits of the training sample the CNN results are not as accurate \citep[see e.g.][]{Yan2020}, we extend the training sample one magnitude fainter than the test sample and the final catalogue to avoid such effects. There are almost 6,000 galaxies that can be simulated to the maximum apparent magnitude of $\mathrm{m}_r < 22.5~\mathrm{mag}$.


\subsection{ETG vs. LTG classification scheme}
\label{sec:ETG_vs_LTG}

In this section, we present our CNN predictions for differentiating between ETGs and LTGs. We denote the ETG as the negative class ($\mathrm{Y}=0$), while the LTG is considered as the positive class ($\mathrm{Y}=1$). The T-Type parameter derived by DS18 is a continuous variable ranging from -3 to 10, where values of T-Type $< 0$ correspond to ETGs and values of T-Type $> 0$ designate LTGs. Unfortunately, the quality of the galaxy images, especially at fainter magnitudes, prevents us from providing such a fine classification for DES galaxies. Separating the sample in two main subclasses (ETGs and LTGs) seems like a reasonable goal for the present catalogue. However, the transition between ETGs and LTGs is smooth and continuous, where intermediate T-Type values are usually assigned to lenticular galaxies (also known as S0s).  Given that this classification is trained in binary mode,  we select a galaxy sample of LTGs and ETGs, therefore, not including intermediate T-Types (-0.5 < T-Type < 0.5). According to this criteria, a total of 1,293 galaxies ($\sim 6 \%$) were excluded. Since the DES observations are deeper than the SDSS ones, the classification based on SDSS imaging could differ for some of the DES galaxies. To improve the quality of our training sample we also excluded 1,488 galaxies ($\sim 7 \%$) with wrong labels in DS18 identified after a visual inspection of the miss-classifications for the predictions of a preliminary model. Then, we re-trained the model without those objects. In summary, our primary training sample consists of 17,132 galaxies with $|\mathrm{T\text{-}Type}| > 0.5$  and accurate spec-$z$, and is magnitude-limited with $\mathrm{m}_r < 17.7$ mag (as the original DS18 catalog).

\subsubsection{Training}

 As described in section~\ref{sec:training}, we use a combination of the primary and the simulated training samples. Nevertheless, from the primary training sample of 17,132 galaxies, we randomly select a subset of 1,132 galaxies and their corresponding \textit{rnd} and \textit{max} simulated samples that we never use for training. We denote this subset as the `test sample' and we use it to check the models' performances. Since none of these galaxies (neither the original nor the simulated ones) have been shown to the CNN, using this subset as a test sample is a secure way to check the results of our classification scheme. Since we want to test our models to $\mathrm{m}_r < 21.5$ we only show results for galaxies up to that apparent magnitude threshold. In figure~\ref{fig:appmag}, we show the apparent magnitude distribution of the test sample to $\mathrm{m}_r < 21.5$. The primary test sample consists of 1,132 galaxies, and the \textit{rnd} and \textit{max} test samples include 1,088 and 623 galaxies, respectively. Therefore, the test sample includes a total 2,843 galaxies to $\mathrm{m}_r < 21.5$, of which 1,557 (55\%) are labelled as ETGs and 1286 (45\%) are labelled as LTGs.

After removing the galaxies belonging to the test sample, we end up with a training sample of 48,000 galaxy images (16,000$\times$3), with roughly 50\% of each class (ETGs and LTGs). We randomly shuffle the galaxies in the training sample and apply a $k$-fold cross-validation technique (with $k=5$) for which the original training sample is randomly partitioned into $k$ equal sized sub-samples. Of the $k$ sub-samples, a single sub-sample is retained as the validation data while training the model, and the remaining $k-1$ sub-samples (38,400) are used as training data. By doing so we ensure that each of the 5 CNN models derived is trained with a different set of images and a different initialisation; this provides a (rough) estimate of the classification uncertainty. 

\subsubsection{Testing}
\label{sec:ETG_vs_LTG-test}

\begin{figure}
\centering
 \includegraphics[width=\columnwidth]{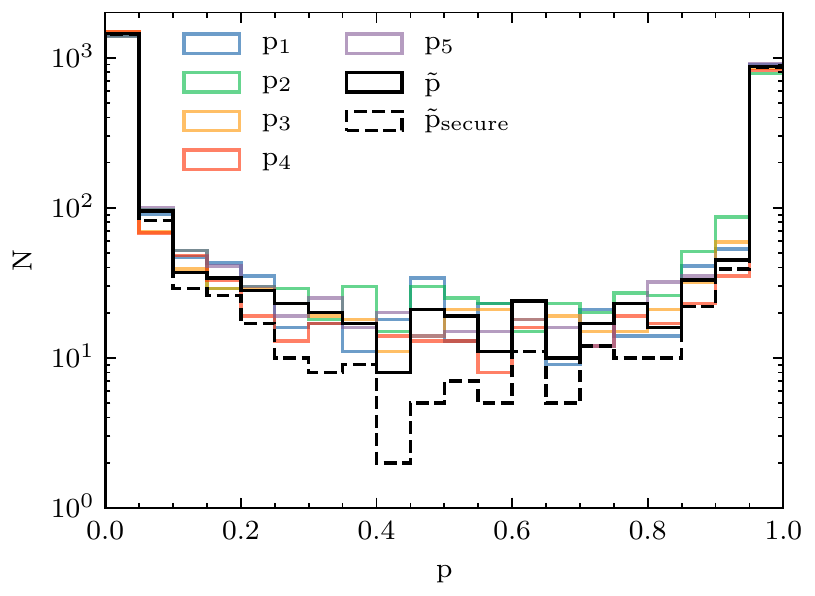}
 \caption{Distribution of the predicted probabilities (p$_i$) for the test sample $i$ in the ETG vs. LTG classification scheme. Black solid histogram corresponds to the distribution of the median probability of the 5 models ($\tilde{\mathrm{p}}$), while the dashed black histogram shows the the distribution of the median probability of the 5 models only for the secure classifications (i.e., those with a $\Delta\mathrm{p}$ < 0.3).
\label{fig:predprob}}
\end{figure}

\begin{figure}
\centering
 \includegraphics[width=\columnwidth]{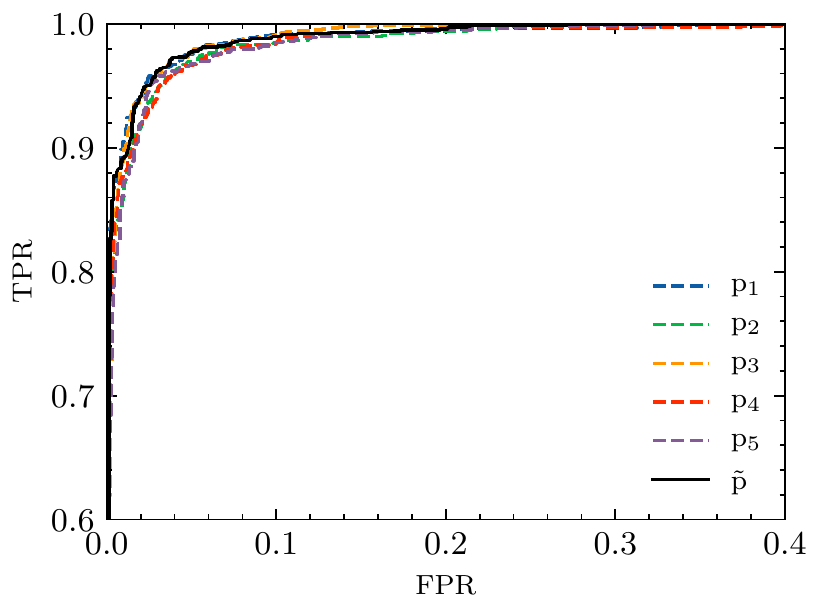}
  \includegraphics[width=0.88\columnwidth]{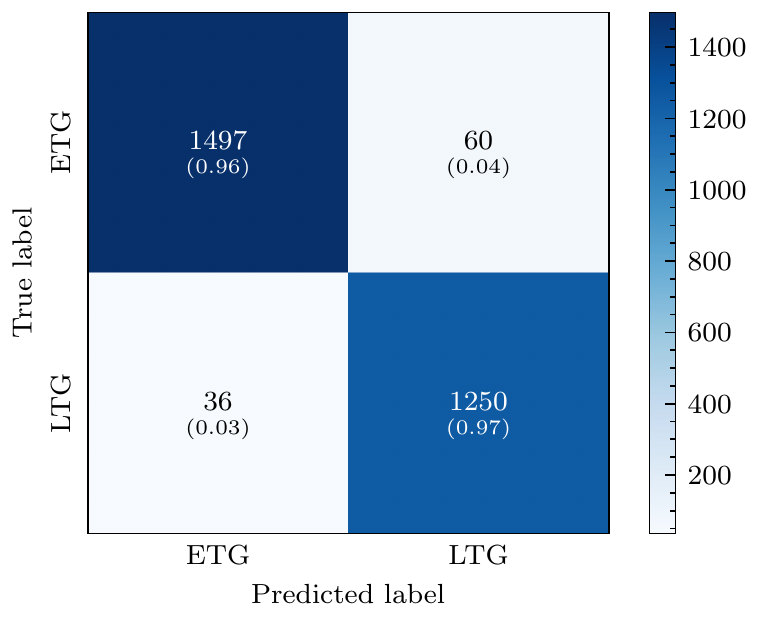}
 \caption{Top panel: ROC curves for the five predicted probabilities (dashed colored lines). Solid black line corresponds to the ROC curve when the predicted probability is equal to the median value ($\tilde{\mathrm{p}}$) of the five predicted probabilities ($\mathrm{p}_i$). Bottom panel: Confusion matrix for the ETG vs. LTG classification scheme. In each cell, we show the number of candidates and the fraction of candidates (in brackets) of TN (top-left), FN (top-right), FP (bottom-left) and TP (bottom-right).
\label{fig:roc}}
\end{figure}

One way to check the reliability of the model predictions made for the ETG vs. LTG classification scheme (hereafter interpreted as a predicted probability, $\mathrm{p}_i$), is to compute the difference in the maximum and the minimum values for the predicted probabilities of the 5 models (expressed as $\Delta\mathrm{p}$). We find that $92.3\%$ of the galaxies from the test sample have $\Delta\mathrm{p} < 0.3$, and we designate these as secure classifications. The remaining $7.7\%$ of the galaxies within the test sample have less secure classifications. In figure~\ref{fig:predprob}, we show the distribution of the predicted probabilities for the test sample along with the distribution of their median probability value (for the full and for the secure classifications). Note that the majority of the insecure classifications are clustered around intermediate values of p. 

As extensively done in literature \citep[see][for instance]{Powers2011}, we also check the accuracy of our models by computing the area under the ROC curve (ROC AUC) for the different predicted probabilities. The ROC curve is a representation of the false positive rate ($\mathrm{FPR = FP/N}$, i.e., the ratio of the number of false positives to negative cases) versus the true positive rate ($\mathrm{TPR = TP/P}$, i.e., the ratio of the number of true positives to positive cases) for different probability thresholds. A good classifier maximises TP and minimises FP values.  The top panel of figure~\ref{fig:roc} shows the ROC curves for the 5 models, and for their median value.  Good classifiers should be as close as possible to the left hand and upper boundaries of this panel, and this is clearly true for all the curves, which all have ROC AUC values of 0.99.

A complementary way to test the model performance is the precision (Prec) and recall (R) scores \citep[e.g.][]{Dieleman2015}, which can be defined as follows:
\begin{align}
    \mathrm{Prec} &= \mathrm{\frac{TP}{TP+FP}};\\
    \mathrm{R} &= \mathrm{\frac{TP}{TP+FN} = TPR},
\end{align}
where the separation between positive and negative samples is determined with respect to a probability threshold. The precision is intuitively the ability of the classifier not to label as positive a sample that is negative (or a purity/contamination indicator). The recall is intuitively the ability of the classifier to find all the positive samples (i.e., a completeness indicator). Additionally, the accuracy of the model prediction is defined as the fraction of correctly classified instances:
\begin{equation}
    \mathrm{Acc = \frac{TP+TN}{P+N}}
\end{equation}
 We derive the probability threshold ($\mathrm{p_{thr}}$) that optimises the ROC curve (i.e., maximises TPR and minimises FPR), but depending on the user purpose, one can vary the $\mathrm{p_{thr}}$ to obtain a more complete or less contaminated sample.

In table~\ref{tb:models}, we present a summary of these different estimators for the 5 independent model predictions ($\mathrm{p_i}$) and for their median value  ($\mathrm{\tilde{p}}$). We have already noted that the ROC AUC equals 0.99 for all the cases:  Prec, R and Acc range from 0.95 to 0.97. Besides, if we assume $\tilde{\mathrm{p}}$ as our fiducial probability, there are only 36 FN and 60 FP in the test sample (3\% and 4\%, respectively), as shown in the confusion matrix in the bottom panel of figure~\ref{fig:roc}, which translates into an Acc$=0.97$. If we restrict attention to the subset of secure classifications, the number of FN and FP decreases to 20 and 28, respectively. This restriction leads to an accuracy classification score for the secure subset of $\mathrm{Acc} = 0.98$. Nevertheless, $\sim 80\%$ of the insecure classifications are still valid (FN and FP are $16+32=48$ out of 219 galaxies that comprise 7.7$\%$ of the test sample).

Even for the subset of secure candidates (92$\%$ of the test sample) there are some galaxies with intermediate probabilities. We define a \textit{robust} sub-sample of ETGs and LTGs as those with $max(\mathrm{p}_i) < 0.3$ and $min(\mathrm{p}_i) > 0.7$, respectively. Note that \textit{robust} classifications are (by definition) within the secure subset. We find that 1,391 galaxies are classified as \textit{robust} ETGs  (they are $53\%$ of the secure and a $49\%$ of the whole test sample). On the other hand, 1,077 galaxies are \textit{robust} LTGs ($41\%$ of the secure and a $38\%$ of the whole test sample). The remaining $6\%$ of the secure sample (156 galaxies) are intermediate (but still secure) candidates. Nevertheless, if we use the median of the probabilities  $\mathrm{\tilde{p}}$ and the optimal threshold of 0.40, most of the galaxies from the secure sample (92$\%$) are still correctly classified. These results demonstrate that the model is able to separate ETGs and LTGs even for the intermediate candidates.

Additionally, we apply our models to the subset of 1,293 galaxies (and their simulated counterparts) with -0.5 < T-Type < 0.5 that we did not include in the training of our models. In total, we classified 3,879 galaxies with intermediate values of T-Type (3 $\times$ 1,293, the original galaxies and the two simulated counterparts), of which 2,004 are ETGs (i.e., -0.5 < T-Type < 0.0) and 1,875 are LTGs (i.e., 0.0 < T-Type < 0.5). We find that $62\%$ of them are secure classifications (i.e., $\Delta\mathrm{p} < 0.3$), significantly lower than the same value for the whole test sample. The number of this subset of galaxies that are classified as ETGs and LTGs is 2,026 ($52\%$ of the total) and 1,853 ($48\%$ of the total), respectively. We also find that 1,348 ($35\%$ of the total) and 874 ($23\%$ of the total) galaxies are classified as \textit{robust} ETGs and \textit{robust} LTGs, respectively. Only 177 galaxies ($5\%$ of the total) are classified as intermediate but still secure candidates. In terms of accuracy, $75\%$ of these galaxies are correctly classified as ETGs, while $88\%$ of these galaxies are correctly classified as LTGs. Therefore, our classifications are reliable (although more uncertain) even for those objects with intermediate values of T-Type (i.e., -0.5 < T-Type < 0.5) that are a-priori difficult to classify.

Note that the fraction of galaxies with intermediate T-Types is very small (6\% in the primary training sample).  Including these galaxies in the test set (assuming the same fraction as in the primary training sample) would reduce the accuracy from 97\% to 96\%. The fraction of such objects in the full DES catalogue (section~\ref{sec:des_morph}) is unknown and their labels are uncertain (see the large scatter in figure 11 from DS18).  As a result, it is difficult to quantify how they impact the overall accuracy.  Therefore, we only quote the final accuracy of the models after such objects have been removed.

\begin{figure}
\centering
 \includegraphics[width=\columnwidth]{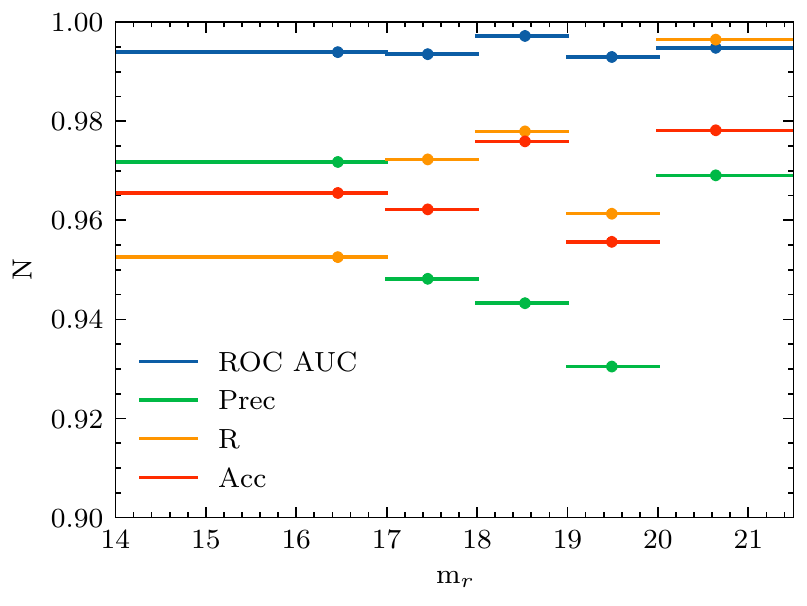}
 \caption{ETG vs. LTG model performance (in terms of ROC AUC, precision, accuracy and recall) as a function of magnitude for the test sample. The values are calculated using the $\mathrm{p_{thr}}$ of the median model $\mathrm{\tilde{p}}$ from table \ref{tb:models}. The lack of dependence of the metrics with magnitude demonstrates that the model is able to correctly classify even the fainter galaxies.  
\label{fig:model-perf}}
\end{figure}

One of the key questions we would like to answer is how much are the results of our classification affected by the galaxy brightness. In figure \ref{fig:model-perf} and in table \ref{tb:models-mag}, we show how the metrics used to test the model performance change with apparent magnitude. In general, there are very small variations, with the AUC ROC being the most stable parameter (> 0.99 always). The accuracy range is also small (0.96 < Acc < 0.98), while the precision and recall show variations of $\sim$5\%. There is no clear trend with apparent magnitude, i.e., the models seem to be able to distinguish between ETGs and LTGs regardless of the faintness of the images. We did the same exercise by dividing the test sample in bins of half-light radius, finding accuracy values above $94\%$, even for the smallest galaxies. Evidently, CNNs detect features hidden to the human eye and therefore classify significantly better than visual inspection.

\begin{table}
\begin{center}
\caption{Summary of the ETG vs. LTG model performance for the five runs: optimal threshold ($\mathrm{p_{thr}}$), area under the ROC curve, precision, recall and accuracy values. The last row shows the values obtained for the median probability of the five runs, $\mathrm{\tilde{p}}$, which we use throughout the paper as the standard model.}
\label{tb:models}
\begin{tabular}{ccccccc}
\hline
\hline
Model & $\mathrm{p_{thr}}$& ROC AUC  & Prec & R & Acc\\
\hline
\hline
$\mathrm{p_1}$ & 0.49 & 0.99 & 0.97 & 0.96 & 0.97\\
$\mathrm{p_2}$ & 0.46 & 0.99 & 0.95 & 0.96 & 0.96\\
$\mathrm{p_3}$ & 0.39 & 0.99 & 0.96 & 0.97 & 0.97\\
$\mathrm{p_4}$ & 0.35 & 0.99 & 0.95 & 0.97 & 0.96\\
$\mathrm{p_5}$ & 0.54 & 0.99 & 0.96 & 0.96 & 0.96\\
\hline
$\mathrm{\tilde{p}}$ & 0.40 & 0.99 & 0.95 & 0.97 & 0.97\\
\hline
\end{tabular}
\end{center}
\end{table}

\begin{table}
\begin{center}
\caption{Summary of the ETG vs. LTG performance in magnitude bins. The values are calculated using the $\mathrm{p_{thr}}= 0.40$ obtained for the full test sample and the median model $\ptilde$ .}
\label{tb:models-mag}
\begin{tabular}{lcccc}
\hline
\hline
Mag bin & ROC AUC  & Prec & R & Acc\\
\hline
\hline
 14 < $\mathrm{m}_{r}$ < 21.5 & 0.99  & 0.95   & 0.97  & 0.97  \\
 \hline
 14 < $\mathrm{m}_{r}$ < 17 & 0.99  & 0.97  & 0.95  & 0.97\\
 17 < $\mathrm{m}_{r}$ < 18 & 0.99  & 0.95  & 0.97   & 0.96 \\
 18 < $\mathrm{m}_{r}$ < 19 & 1.00  & 0.94  & 0.98  & 0.98 \\
 19 < $\mathrm{m}_{r}$ < 20 & 0.99  & 0.93  & 0.96   & 0.96 \\
 20 < $\mathrm{m}_{r}$ < 21.5 & 0.99 & 0.97  & 1.00   & 0.98 \\
\end{tabular}
\end{center}
\end{table}


\subsection{Face-on vs. Edge-on classification scheme}
\label{sec:edgeon}

In this section, we present our CNN predictions for the second classification scheme to distinguish face-on vs. edge-on galaxies. Whereas what we mean by `edge-on' is intuitively obvious, and we treat these as the positive class ($\mathrm{Y}=1$), we use the term `face-on' to refer to the objects that are not edge-on (i.e., the negative class, $\mathrm{Y}=0$), i.e., this classification does not aim to return a continuous output, such as galaxy inclination or ellipticity, but rather to select galaxies that are clearly viewed edge-on. 

\begin{figure}
\centering
 \includegraphics[width=\columnwidth]{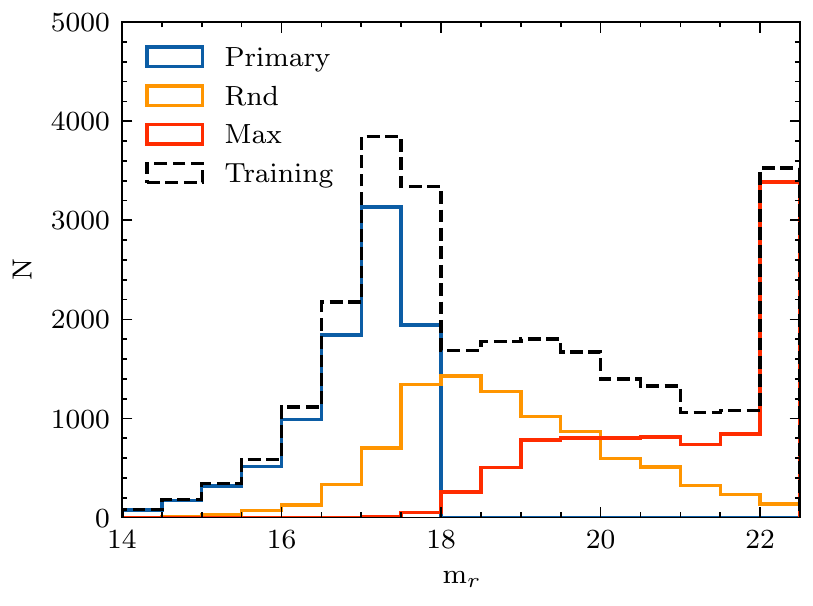}
 \includegraphics[width=\columnwidth]{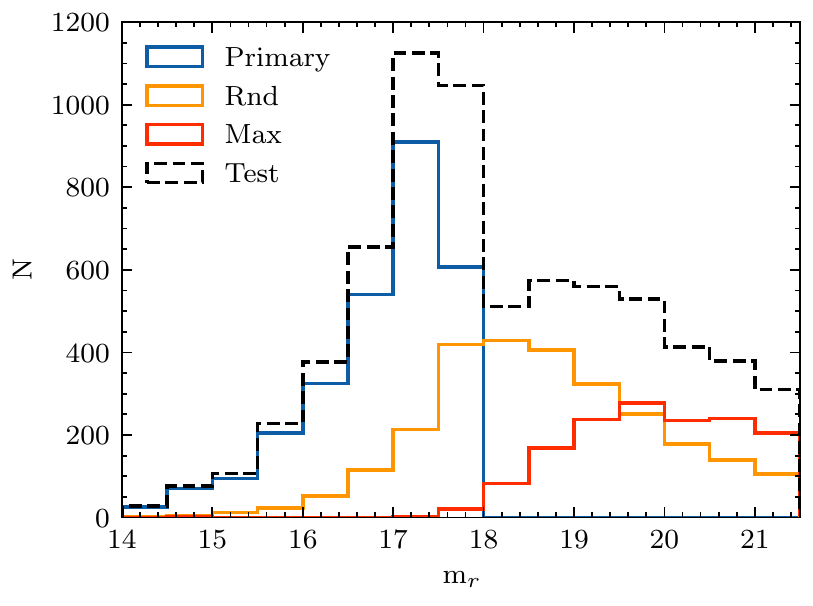}
 \caption{Top panel: Same as figure~\ref{fig:appmag} but for the face-on vs. edge-on classification scheme. Bottom panel: Same as top but only for the test sample (see section~\ref{sec:edgeon}). Note that the CNN predictions are trained with a sample that extends up to $\mathrm{m}_r < 22.5$, while we test the model predictions only up to $\mathrm{m}_r < 21.5$ (the limit of the DES catalogue presented in this work).
\label{fig:appmag_edgeon}}
\end{figure}

\subsubsection{Training}
As for the first classification scheme, our training sample is a combination of the original and the simulated samples. We use the information provided by DS18 on the probability of being  edge-on (see section~\ref{sec:ds18})  to select a reliable sample of galaxies with which to train our CNNs. We define face-on galaxies as those with $\mathrm{p_{edge-on}} < 0.1$ and edge-on galaxies as those with $\mathrm{p_{edge-on}} > 0.9$ , corresponding to 11,783 galaxies. We randomly select 2,783 galaxies (and their simulated versions) as the test sample and the remaining 9,000 for the training sample. The training sample consists of 27,000 galaxies (3$\times$9,000, the originals and their simulated versions) with 23,424 (87\%) face-on galaxies and 3,576 (13\%) edge-on galaxies. As for the ETG vs. LTG model, we train 5 different models with  $k$-folding. We have reserved a total of 8,349 original and simulated galaxies for testing. However, as for the first classification scheme (section~\ref{sec:ETG_vs_LTG}), we only show results for galaxies to $\mathrm{m}_r < 21.5$:  all the 2,783 galaxies within the primary test sample, 2,673 galaxies from the \textit{rnd} test sample and 1,477 galaxies from the \textit{max} test sample. Therefore, the test sample includes a total 6,933 galaxies, of which 6,066 (87\%) are face-on and 876 (13\%) are edge-on. In figure~\ref{fig:appmag_edgeon}, we show the distribution of ($\mathrm{m}_r$) for the different datasets that make up the training and the test samples for the face-on vs. edge-on classification scheme.

Since the fractions of face-on and edge-on galaxies are so unequal, we use balanced weights during the training phase of our CNN. In other words, the algorithm compensates for the lack of examples of one class by dividing the loss of each example by the fraction of objects of that particular class.


\begin{figure}
\centering
 \includegraphics[width=\columnwidth]{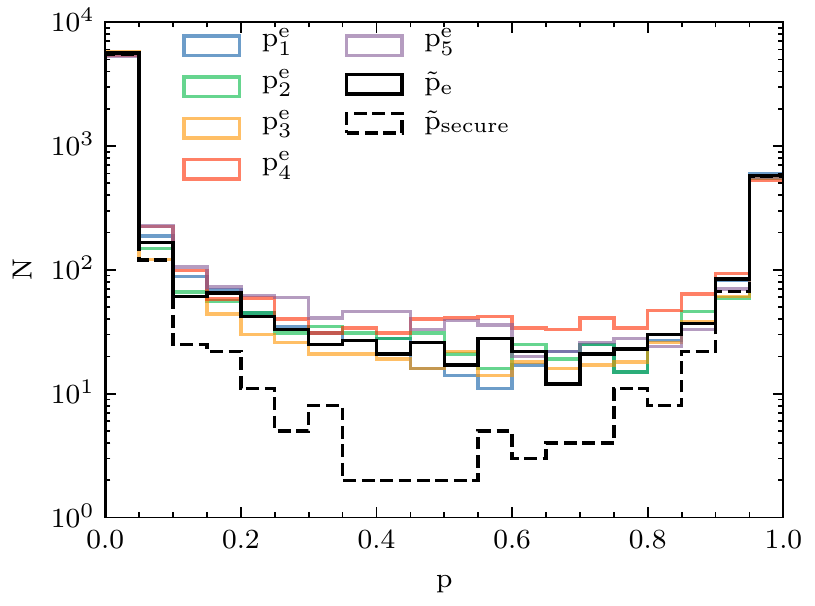}
 \caption{Distribution of the predicted probabilities ($\mathrm{p^e_i}$) for the test sample $i$ in the face-on vs. edge-on classification scheme. Black solid histogram corresponds to the distribution of the median probability of the 5 models ($\petilde$), while the dashed black histogram shows the the distribution of the median probability of the 5 models only for the secure classifications (i.e., those with a $\Delta\mathrm{p_e}$ < 0.3).
\label{fig:predprob_edgeon}}
\end{figure}

\begin{figure}
\centering
 \includegraphics[width=\columnwidth]{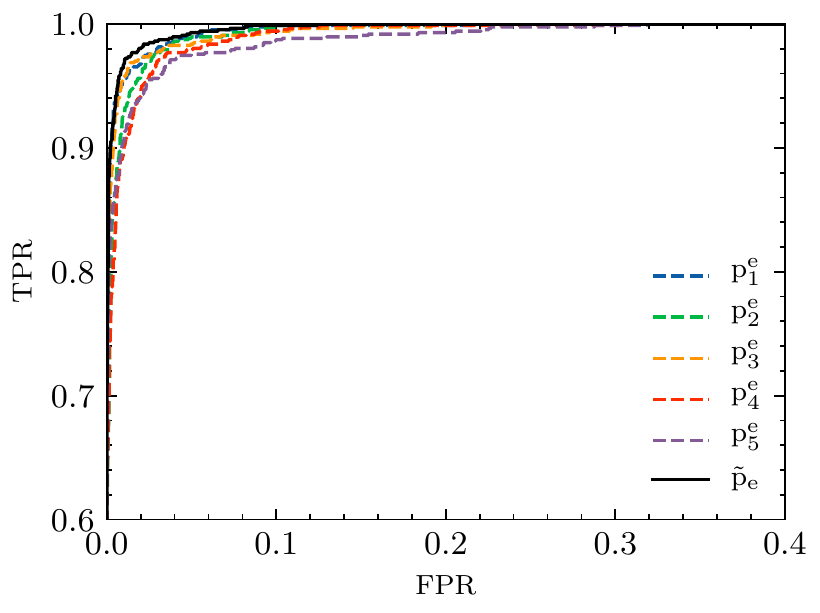}
  \includegraphics[width=0.88\columnwidth]{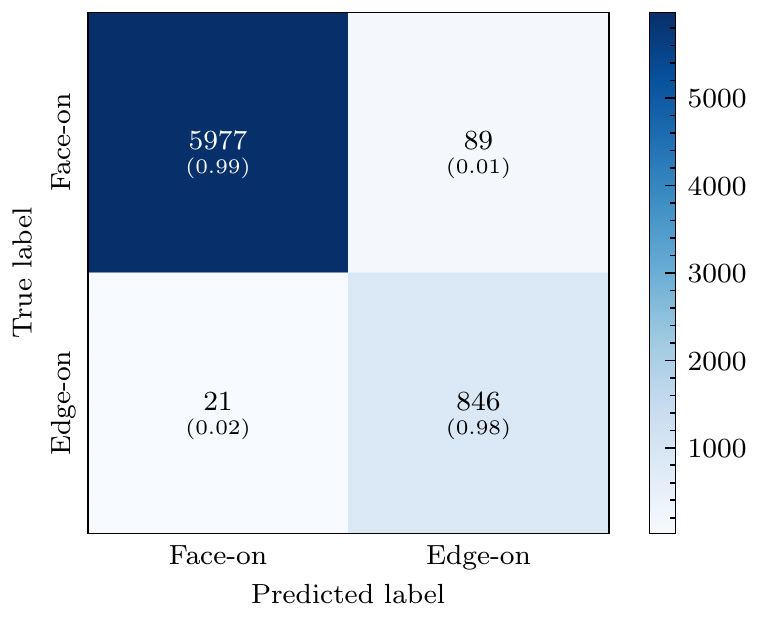}
 \caption{Same as figure~\ref{fig:roc} but for the face-on vs. edge-on classification. Note the better performance of the average model $\petilde$. 
\label{fig:conf_matrix}}
\end{figure}

\subsubsection{Testing}
\label{sec:edgeon-test}

As described in section~\ref{sec:ETG_vs_LTG-test}, we check the accuracy of our model predictions by means of the ROC AUC, Prec, R and Acc estimators. In table~\ref{tb:models-edegon}, we show these values for the 5 face-on vs. edge-on models (denoted as $\mathrm{p^e_i}$) and the median one (denoted as $\petilde$). The top panel of figure~\ref{fig:conf_matrix} shows the ROC curve for the different models while  the bottom panel  summarises the results of $\petilde$  in a confusion matrix showing the number of TN (5977), FP (89), FN (21) and TP (846) along with their respective fractions within the two classes. The median model $\petilde$ model has a better performance than the 5 individual models with Acc$=0.98$ and R$=0.98$, while Prec$=0.90$ is slightly smaller. This is in part due to the unbalanced test sample: the total number of FP is about 1/10 of the TP, although the FP are only $\sim$ 1--2\% of the face-on galaxies. On the other hand, the number of FP (only 89, or $1\%$ of the predicted edge-on galaxies) is considerably lower than the number of TP, which translates into an excellent R value. 

Analogously to the ETG vs. LTG model, we define a  \textit{secure} sub-sample of galaxies for the edge-on classification where  $\Delta\mathrm{p_e} < 0.3$. There are 93\% of \textit{secure} galaxies in the test sample. The \textit{robust} edge-on are galaxies with $min(\mathrm{p}^{e}_i) > 0.7$. We find 668 galaxies classified as \textit{robust} edge-on ($10\%$ of the secure sample and $12\%$ of the whole test sample).

\begin{figure}
\centering
 \includegraphics[width=\columnwidth]{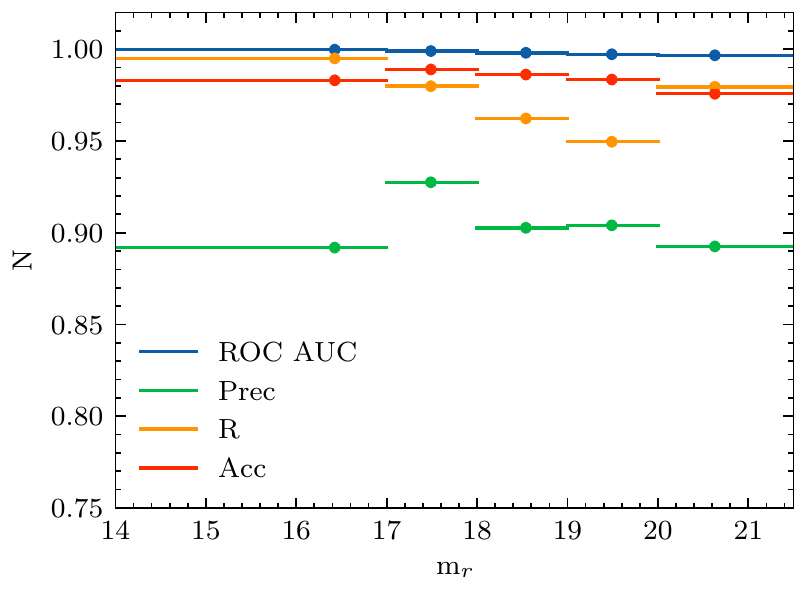}
 \caption{Same as figure~\ref{fig:model-perf} but for the edge-on model performance. The lack of dependence of the metrics with magnitude demonstrates that the model is able to correctly classify even the fainter galaxies
\label{fig:model-perf_edgeon}}
\end{figure}

The dependence of the edge-on classification with apparent magnitude is highlighted in  figure \ref{fig:model-perf_edgeon}, which plots the performance of the $\petilde$ model in the same magnitude bins (summarised in table \ref{tb:models-edegon-mag}). There is a very small variation with apparent magnitude:  the most affected quantity decreases from 0.99 at $14.0 < \mathrm{m}_r < 17.0$ to 0.95 at $19.0 < \mathrm{m}_r < 20.0$. In the same table we show the values obtained for a balanced test sample, robust against class representation. In this case, the precision values are significantly improved (from Prec $=0.90$ to Prec $=0.98$ for the full test sample) while the other indicators are almost unchanged. We did the same exercise by dividing the test sample in bins of half-light radius, finding accuracy values above $96\%$, even for the smallest galaxies.

\begin{table}
\begin{center}
\caption{Summary of the edge-on versus face-on model performance for the five runs: optimal threshold ($\mathrm{p_{thr}}$), area under the ROC curve, precision, recall and accuracy values. The last row shows the values obtained for the median probability of the five runs, $\petilde$, which we use throughout the paper as the standard model.}
\label{tb:models-edegon}
\begin{tabular}{ccccccc}
\hline
\hline
Model & $\mathrm{p_{thr}}$& ROC AUC  & Prec & R & Acc\\
\hline
\hline
$\mathrm{p^e_1}$ & 0.26 & 1.00 & 0.86 & 0.97 & 0.98\\
$\mathrm{p^e_2}$ & 0.21 & 1.00 & 0.83 & 0.98 & 0.97\\
$\mathrm{p^e_3}$ & 0.32 & 1.00 & 0.92 & 0.97 & 0.98\\
$\mathrm{p^e_4}$ & 0.35 & 1.00 & 0.80 & 0.98 & 0.97\\
$\mathrm{p^e_5}$ & 0.29 & 0.99 & 0.79 & 0.97 & 0.96\\
\hline
$\petilde$ & 0.33 & 1.00 & 0.90 & 0.98 & 0.98\\
\hline
\end{tabular}
\end{center}
\end{table}

\begin{table}
\begin{center}
\caption{Summary of the edge-on model performance in magnitude bins. The values are calculated using the $\mathrm{p_{thr}}= 0.33$ obtained for the full test sample and the median model $\petilde$ (in brackets  for a balanced test sample).}
\label{tb:models-edegon-mag}
\begin{tabular}{lcccc}
\hline
\hline
Mag bin & ROC AUC  & Prec & R & Acc\\
\hline
\hline
 14 < $\mathrm{m}_{r}$ < 21.5 & 1.00 (1.00) & 0.90 (0.98)  & 0.98 (0.98) & 0.98 (0.98) \\
 \hline
 14 < $\mathrm{m}_{r}$ < 17 & 1.00 (1.00) & 0.89 (0.97) & 0.99 (0.99) & 0.98 (0.98)\\
 17 < $\mathrm{m}_{r}$ < 18 & 1.00 (1.00) & 0.93 (1.00) & 0.98 (0.98)  & 0.99 (0.99)\\
 18 < $\mathrm{m}_{r}$ < 19 & 1.00 (1.00) & 0.90 (0.99) & 0.96  (0.96) & 0.99 (0.98)\\
 19 < $\mathrm{m}_{r}$ < 20 & 1.00 (1.00) & 0.90 (0.97) & 0.95  (0.95) & 0.98 (0.96)\\
 20 < $\mathrm{m}_{r}$ < 21.5 & 1.00  (1.00)& 0.89  (0.99) & 0.98  (0.98) & 0.98 (0.98)\\
\end{tabular}
\end{center}
\end{table}

\begin{table*}
\begin{center}
\caption{Comparison of the test samples discussed in section \ref{sec:results}. Columns show the total number of galaxies and the corresponding fraction of secure ones ($\Delta\mathrm{p} < 0.3$). Also given are the fractions of the secure galaxies classified as (\textit{robust}) ETGs, LTGs  and edge-on. The last column contains the fraction of \textit{robust} ETGs that are also classified as \textit{robust} edge-on. These cases should be taken with care since only discs should be edge-on.}
\label{tab:results}
\begin{tabular}{lrcccccc}
\hline
\hline
Sample & \# galaxies & Secure ($\Delta\mathrm{p} < 0.3$) & ETGs (\textit{robust})  & LTGs (\textit{robust}) & Secure ($\Delta\mathrm{p_e} < 0.3$) & Edge-on (\textit{robust}) & ETGs$+$edge-on\\
 &    & \% from total & \%  from secure & \%  from secure & \% from total &\%  from secure & \%  from \textit{robust} ETGs \\ 
\hline
\hline
DES DR1 & 26,971,945 & 87 & 12 (10) & 88 (85)  & 73 & 9 (6) & 0.3 \\
DES struct. param. & 6,060,018 & 89 & 9 (7)  & 91 (88) & 79 & 7 (6) & 0.3\\
DES stellar mass & 137,956 & 83 & 48 (44) & 52 (47) & 86 & 7 (6) & 0.5\\
VIPERS & 7,384 & 81 & 22 (20) & 78 (76) & 77 &  2 (2) & 0.1 \\
\hline
\end{tabular}
\end{center}
\end{table*}

The  face-on vs. edge-on classification is useful for different scientific purposes (see section~\ref{sec:intro}), but  might also help as an additional test for the ETG vs. LTG classification  presented in section~\ref{sec:ETG_vs_LTG}. Since only discs can be seen edge-on, a galaxy should not be classified simultaneously as an ETG and edge-on. We find 91 (predicted)  ETGs in the ETG vs. LTG test sample with  $\petilde > 0.33$, corresponding to $\sim$3\% of the test sample. This fraction is reduced to 0.7\% when only \textit{robust} ETG and \textit{robust} edge-on are considered. This small fraction  reassures us about the performance of  the two models. A visual inspection of these galaxies confirms that most of them look like edge-on lenticulars, with a clear bulge and disc but no signs of spiral arms.  This is especially evident for \textit{robust} edge-on ETGs (see figure~\ref{fig:cutouts}). Thus, including the additional information provided by the edge-on vs. face-on classification helps to increase the purity of the ETG sample and is an efficient way to identify edge-on lenticulars.


\section{DES DR1 morphological catalog}
\label{sec:des_morph}

\begin{figure}
\centering
 \includegraphics[width=\columnwidth]{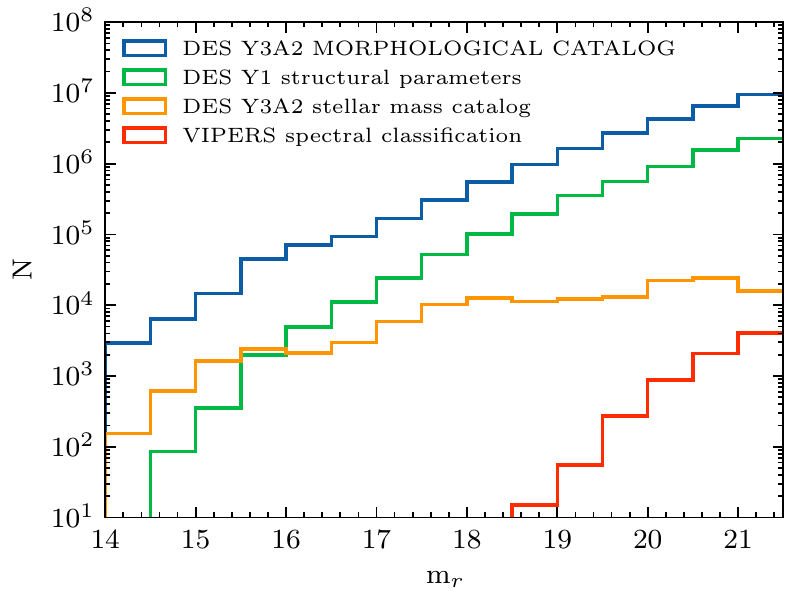}
 \caption{Normalized apparent magnitude distribution for the full DES morphological catalogue presented in this work, as well as  for the three catalogues used for comparison.
\label{fig:mag-distr}}
\end{figure}

In this section, we present the results of applying the classification schemes described in sections~\ref{sec:ETG_vs_LTG} and~\ref{sec:edgeon} to the DES DR1 galaxy catalog presented in section~\ref{sec:desy3}. We briefly summarise the overall results here but address a more exhaustive comparison with other observed galaxy properties in section~\ref{sec:results}. Table \ref{tab:results} summarises the statistics for the full DES morphological catalogue, as well as for three comparison samples, while the magnitude distribution of each sub-sample is shown in figure \ref{fig:mag-distr}. In table~\ref{tab:catalog}, we describe the content of the full DES DR1 morphological catalogue, which will be released along with the paper. Examples of each class at different magnitudes are shown in appendix \ref{appendix}.

For the ETG vs. LTG classification scheme, $87\%$ of the 26,971,945 galaxies in the DES DR1 morphological catalogue are secure classifications, i.e., $\Delta\mathrm{p} < 0.3$ (where $\Delta\mathrm{p}$ corresponds to the maximum difference between the five predicted probabilities). Within this subset of secure classifications, $10\%$ of the galaxies are classified as \textit{robust} ETGs (i.e., $max(\mathrm{p_i}) < 0.3$), while $85\%$ are classified as \textit{robust} LTGs (i.e., $min(\mathrm{p_i}) > 0.7$) . The remaining $5\%$ of the galaxies may be considered as intermediate (but still secure) candidates. Being less conservative, $\sim 12\%$ of the galaxies  from the subset of secure classifications are classified as ETGs (i.e., $\mathrm{\tilde{p}} < \mathrm{p_{thr}} = 0.4$) while, consequently, $\sim 88\%$ are classified as LTGs (i.e., $\mathrm{\tilde{p}} > \mathrm{p_{thr}} = 0.4$). The much larger fraction of LTGs with respect to ETGs in a magnitude limited sample is consistent with  previous work (see e.g., \citealt{Pozzetti2010}).

\begin{figure}
\centering
 \includegraphics[width=\columnwidth]{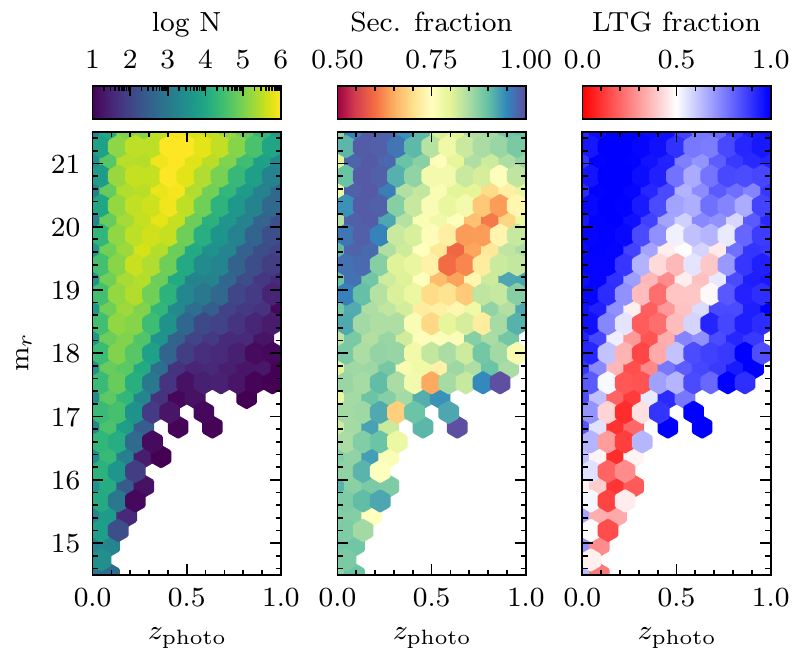}
 \caption{Apparent magnitude ($\mathrm{m}_r$) vs. photometric redshift ($z_\mathrm{photo}$) for the secure subset ($\Delta_\mathrm{p} < 0.3$) within the DES DR1 morphological catalog. Left-hand panel shows the number of galaxies (in log-scale) in each hexagonal bin.  Middle panel shows the fraction of secure galaxies. Right-hand panel indicates the predicted ETG/LTG fraction. A predicted LTG fraction of 1 means that $100\%$ of the galaxies in a particular hexagonal bin are LTGs, while a predicted LTG fraction of 0 indicates $100\%$ of the objects in the bin are ETGs. The brightest galaxies at low $z_\mathrm{photo}$ are dominated by ETGs, while the faint galaxies are predominantly LTGs. The fraction of secure classified galaxies is relatively constant for the (observed) bright galaxies and the fraction of insecure galaxies increases with  $z_\mathrm{photo}$. The average fraction of secure galaxies is 87\%  and  greater than 50\% even in the more uncertain regions of the $\mathrm{m}_r$-$z_\mathrm{photo}$ plane.
\label{fig:appmag_vs_photoz}}
\end{figure}

Figure~\ref{fig:appmag_vs_photoz} shows how the galaxies  the whole DES DR1 morphological catalog populate the apparent magnitude ($\mathrm{m}_r$) and photometric redshift ($z_\mathrm{photo}$) plane, color coded by density, secure fraction and  predicted LTG fraction. The predicted LTG fraction is computed as the average of the predicted labels (i.e., 0 for ETGs, 1 for LTGs) of the galaxies in each bin. As expected, the brightest galaxies at low $z_\mathrm{photo}$ are dominated by ETGs, while the faint galaxies are predominantly LTGs. The fraction of secure classified galaxies is relatively constant for the (observed) bright galaxies and there is an interesting trend with redshift for galaxies fainter than $\mathrm{m}_r$ > 19: the fraction of insecure galaxies increases with  $z_\mathrm{photo}$, as expected. In any case, note that the average fraction of secure galaxies is 87\%; it remains greater than 50\% even in the more uncertain regions of the $\mathrm{m}_r$-$z_\mathrm{photo}$ plane.

Although most of the faintest (observed) galaxies are classified as LTGs, the classification model is able to retrieve a significant fraction of ETGs ($\sim 50\%$) at intermediate $z_\mathrm{photo} \sim 0.5$ and $\mathrm{m}_r \gtrsim 20.0~\mathrm{mag}$. Note that the faint, low redshift population corresponds to intrinsically faint (and therefore low mass) galaxies, that are, in general, LTGs. Unfortunately, there are no additional parameters with which to further test the full DES DR1 catalogue. We cross-correlate this sample with other available measurements in the following sections.

For the edge-on vs. face-on classification scheme, $18\%$ of the galaxies have values of $\petilde > 0.33$ ($9\%$ if the limit is $min(\mathrm{p^e_i}) > 0.70$). The fraction of \textit{robust} ETGs with $\petilde > 0.33$ is less than $\sim 3\%$ (0.3$\%$ if $min(\mathrm{p^e_i}) > 0.70$). This small fraction is reassuring since, as explained in section \ref{sec:edgeon-test}, edge-on galaxies should only be discs (and therefore LTGs). We strongly recommend that users combine the two classifications since many of these galaxies could actually be miss-classified LTGs or edge-on lenticulars. Some examples are shown in figure \ref{fig:cutouts}.

\begin{table*}
\begin{center}
\caption{Content of the full DES DR1 morphological catalogue.}

\label{tab:catalog}
\begin{tabular}{ll}
\hline
\hline
COADD\_OBJECT\_ID & Unique object ID for Y3 coadd processing \\
RA &  Right ascension  (J2000.0 in degrees) \\
DEC &  Declination (J2000.0 in degrees) \\
MAG\_AUTO\_R  & Apparent magnitude in an elliptical aperture shaped by the Kron radius ($\mathrm{m}_r$ throughout the paper) \\ 
FLUX\_RADIUS\_R & Radius (in pixels) of the circle containing half of the flux of the object ($\mathrm{r}_r$ throughout the paper) \\
$\mathrm{P}i\_\mathrm{LTG}$ & Probability of being LTG for each of the 5 models (with $i = [1,5]$) \\
$\mathrm{MP}\_\mathrm{LTG}$ & Median probability of the 5 models of being LTG \\
$\mathrm{P}i\_\mathrm{EdgeOn}$  & Probability of being edge-on for each of the 5 models (with $i = [1,5]$) \\
$\mathrm{MP}\_\mathrm{EdgeOn}$  & Median probability of the 5 models of being edge-on \\
FLAG\_LTG & Classification for ETG vs. LTG model; 0=ETG, 2=secure ETG, 4=robust ETG; 1=LTG, 3=secure LTG, 5=robust LTG\\
FLAG\_EdgeOn & Classification for edge-on model; 0= no edge-on, 1=edge-on, 2= secure edge-on, 3=robust edge-on\\
\hline
\end{tabular}
\end{center}
\end{table*}


\section{Validation of the classification on real DES galaxies}
\label{sec:results}


Although the results presented in sections \ref{sec:ETG_vs_LTG-test} and  \ref{sec:edgeon-test} show the CNNs perform well, it may be argued that the tests were done on a similar set of simulated images as the ones used to train the CNNs.  To further test the goodness of the morphological classification on real DES DR1 galaxy images we now present a comparison with other available data (both photometric and spectroscopic). 


\subsection{DES DR1 stellar mass catalog}\label{sec:palmese}

The DES DR1 stellar mass catalog is the result of running the \texttt{LePhare} code \citep{Arnouts2011} on the DES DR1 galaxy catalog using \citet{Bruzual2003} templates, three different metallicities (including solar), Chabrier IMF and exponentially declining star formation histories \citep[similarly to][]{Palmese2020}. The redshift of each galaxy is assumed to be equal to the mean photo-$z$ mean statistic obtained from  multi-object fitting photometry \citep[MOF,][]{Drlica-Wagner2018}. The resulting catalog  contains estimates of the stellar mass and the absolute magnitude for $\sim$184 million galaxies.

\begin{figure}
\centering
 \includegraphics[width=\columnwidth]{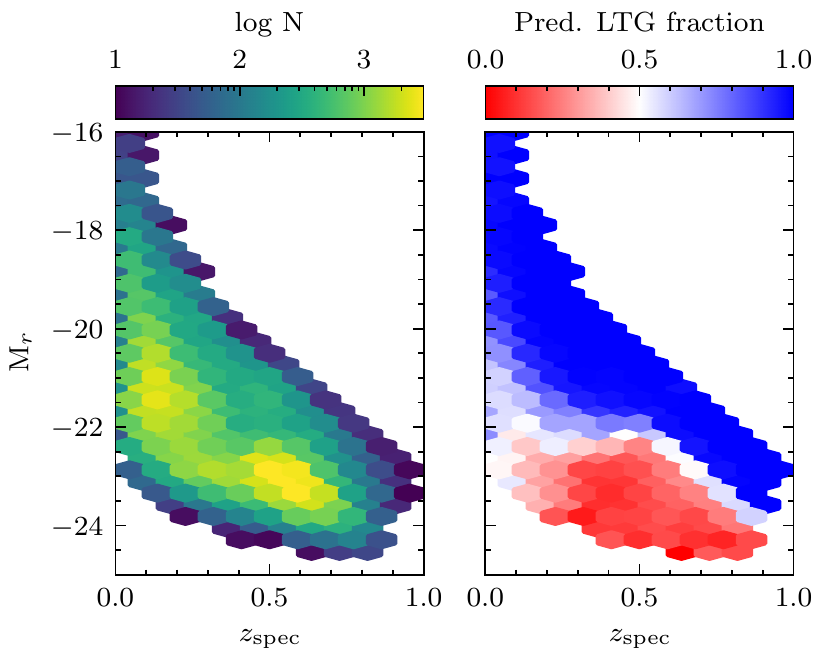}
 \caption{Absolute magnitude ($\mathrm{M}_r$) vs. spectroscopic redshift ($z_{\mathrm{spec}}$) for the secure subset ($\Delta_\mathrm{p} < 0.3$) for the comparison sample of the DES DR1 stellar mass catalog. Left-hand panel shows the number of galaxies (in log-scale) in each bin. Right-hand panel indicates the predicted ETG/LTG fraction. There is a clear separation between the ETG and LTG populations, with the (intrinsically) brightest galaxies at each redshift dominated by the ETGs, as expected. The lack of ETGs at the lowest redshifts ($z \lesssim 0.2$) is due to the scarcity of massive ETGs in such a small volume.
\label{fig:absmag_vs_appmag}}
\end{figure}

To select galaxies in the DES DR1 stellar mass catalog for which the stellar mass and absolute magnitude estimates are reliable (given the large uncertainties associated with the $z_{\mathrm{photo}}$), we cross-match the above mentioned catalog with a spectroscopic compilation from several surveys \footnote{J. Gschwend, private communication (see also \citealt{Gschwend2018A}).} and select galaxies with $(|z_{\mathrm{photo}}-z_{\mathrm{spec}}|)/(1+z_{\mathrm{spec}}) < 0.05$. The sample used for the comparison with our work consists of 137,956 galaxies covering a redshift range of $0 < z < 1$ and the magnitude  shown in figure \ref{fig:mag-distr}. The summary of the statistics is shown in table \ref{tab:results}.

For this sub-sample,  $86\%$ of the galaxies show secure classifications (i.e., $\Delta \mathrm{p} < 0.30$). The fraction of \textit{robust} ETGs and LTGs is 44 and 47\%, i.e., this is a much more balanced sample compared to the full DES DR1 (and also to the other sub-samples shown in table  \ref{tab:results}).  We note that the magnitude distribution of this subset of galaxies is relatively flat, meaning that a large fraction of the faint LTGs may be missing. Regarding the second classification, 7\% of the galaxies are edge-on and less that 0.5\% of the \textit{robust} ETGs are classified as \textit{robust} edge-on.

In figure~\ref{fig:absmag_vs_appmag}, we show how the galaxies populate the absolute  magnitude -- redshift plane ($\mathrm{M}_r$ and $z_{spec}$, respectively). There is a clear separation between the ETG and LTG populations, with the (intrinsically) brightest galaxies at each redshift dominated by the ETGs, as expected. The lack of ETGs at the lowest redshifts ($z \lesssim 0.2$) is due to the scarcity of massive ETGs in such a small volume.
 

\subsection{DES Y1 structural parameters}
\label{sec:tarsitano}

The DES Y1 structural and morphological catalogue presented in \citet{Tarsitano2018} consists of $\sim$50 million objects selected from the first year of the DES. For a comparison with our predicted morphologies, we use the single S\'ersic index ($\mathrm{n}_r$) and the ellipticity ($\epsilon_r$) obtained with GALFIT for the \textit{r}-band. Following Appendix B3.2 of \citet{Tarsitano2018}, we extract a clean sample of validated and calibrated objects by applying the recommended cuts $\mathrm{FIT\_STATUS\_R=1}$ and $\mathrm{SN\_R>10}$ in the \textit{r}-band. We also select objects with realistic values for the S\'ersic index and ellipticity within $0 < \mathrm{n}_r < 10$ and $0 < \epsilon_r < 1$, respectively. These criteria are fulfilled by a 54\% of the objects in the catalogue. Then, we cross-match the resulting catalog with our DES DR1 catalog to $\mathrm{m}_r < 21.5~\mathrm{mag}$ and excluded ($\sim 600$) objects with unreliable redshifts (i.e., $z_\mathrm{photo} < 0.0$).  Finally, we construct a catalog for comparison with 6,060,018 ($\sim12\%$ of the original catalogue) galaxies for which accurate $\mathrm{n}_r$, $\epsilon_r$ and apparent magnitudes are available. The magnitude distribution is shown in figure \ref{fig:mag-distr} and the median of the apparent magnitude of the selected sub-sample is $\mathrm{\tilde{m}_r} = 20.8~\mathrm{mag}$.

As detailed in table \ref{tab:results}, $89\%$ of the galaxies in this subset show secure ETG/LTG classifications (i.e., $\Delta \mathrm{p} < 0.30$). While the fraction of edge-on, 7\%,  is very similar to the other sub-samples, the fraction of \textit{robust} ETGs and LTGs (7 and 88 \%, respectively) is very uneven. The \textit{r}-band magnitude distribution is similar to the DES DR1 morphological catalogue, although missing some galaxies at the very bright end), which can explain the larger fraction of LTG for this subset.

\begin{figure*}
\centering
 \includegraphics[width=\columnwidth]{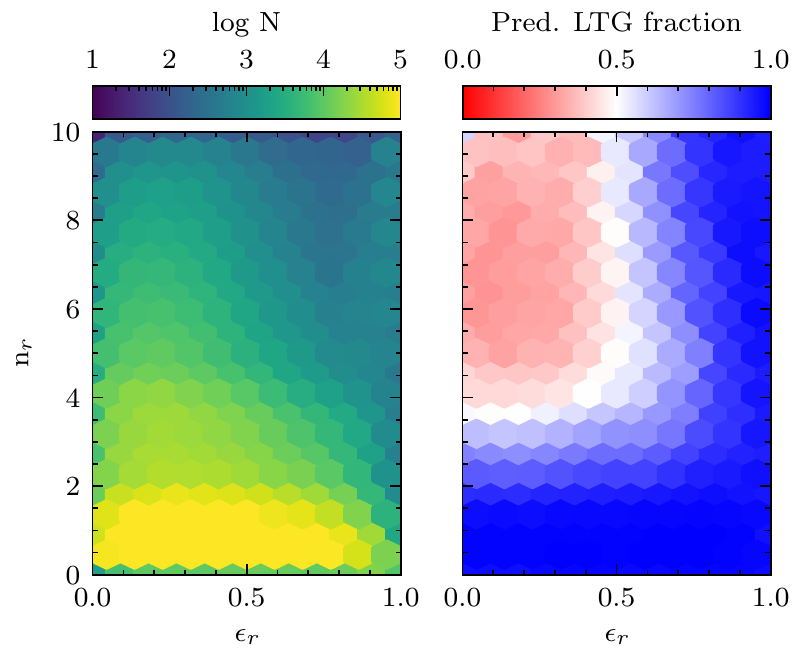}
 \includegraphics[width=\columnwidth]{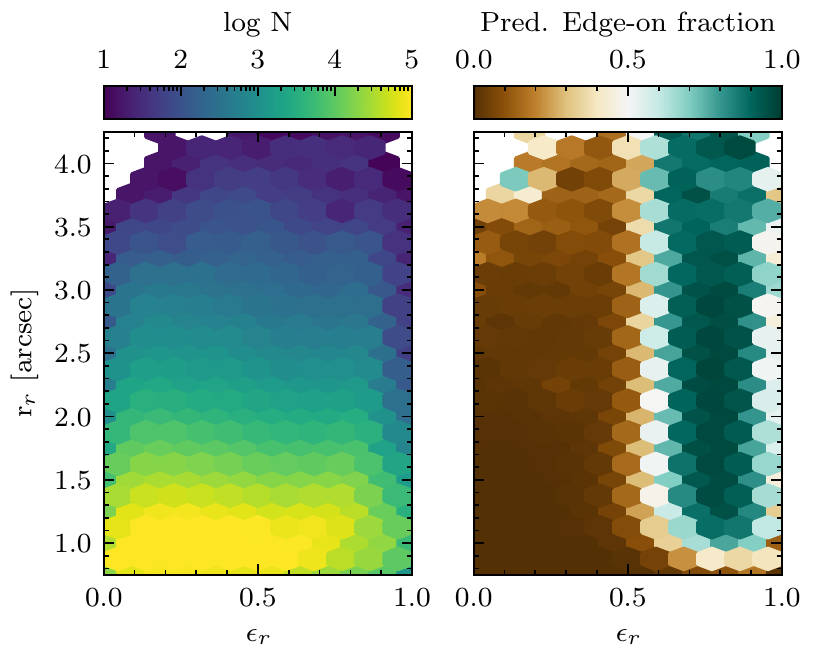}
 \includegraphics[width=\columnwidth]{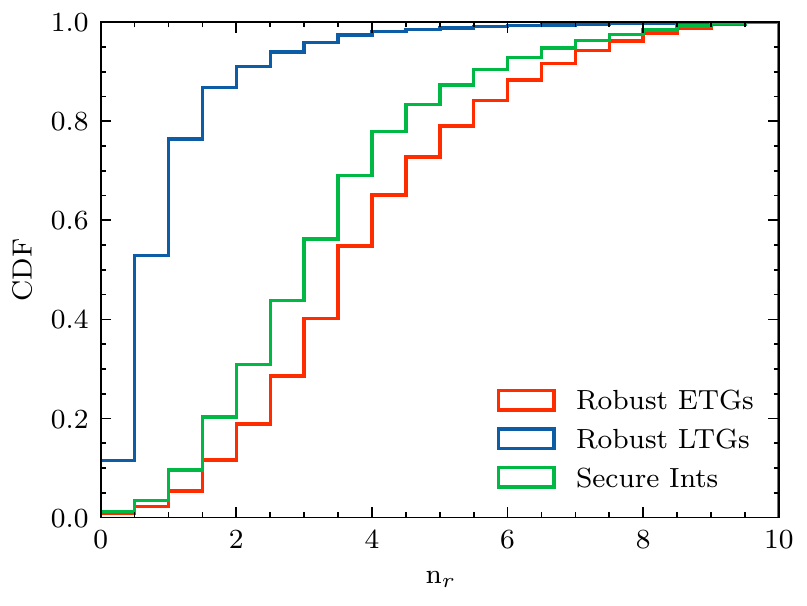}
 \includegraphics[width=\columnwidth]{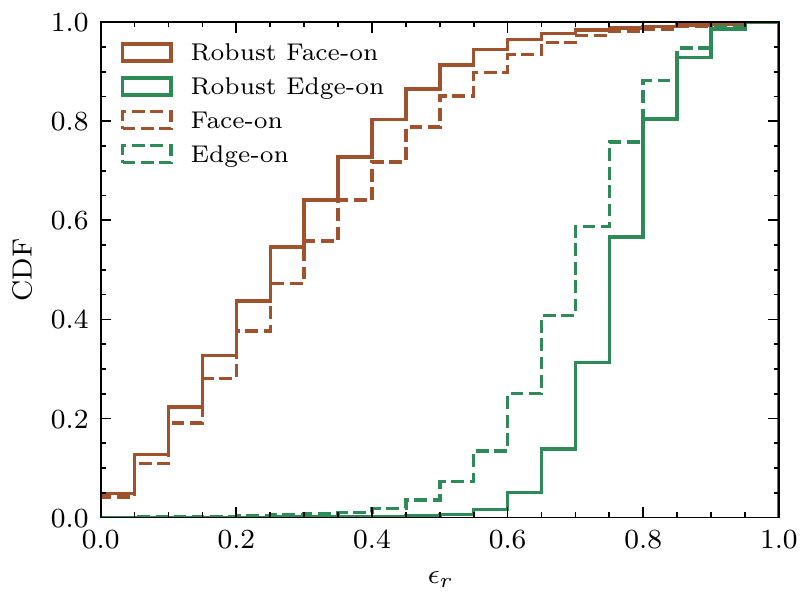}
 \caption{Top-left panels:  S\'ersic index vs. ellipticity for the sample in common with the DES Y1 structural parameters catalogue. The bins are color coded by number density and fraction of LTG over the total. Bottom-left panel: Cumulative distribution function (CDF) of the S\'ersic index for  the ETG vs. LTG classification scheme: red histogram corresponds to the \textit{robust} ETGs, those with $max (\mathrm{p_i}) < 0.30$; blue histogram shows the \textit{robust} LTGs, those with $min (\mathrm{p_i}) > 0.70$; green histogram corresponds to the intermediate but secure candidates. Top-right panels: Observed radius vs. ellipticity  color coded by number density and fraction of edge-on galaxies over the total. Bottom-right panel: CDFs for the ellipticity. Black histograms show the CDFs for the face-on galaxies with $\petilde < 0.33$ (dashed) and $max (\mathrm{p^e_i}) < 0.30$ (solid). Orange histograms correspond to the CDFs for the edge-on galaxies with $\petilde > 0.33$ (dashed) and  $min (\mathrm{p^e_i}) > 0.70$ (solid). The very different distributions of the sub-samples is an indicator of the accuracy of our model predictions in real DES galaxy images.
\label{fig:cum_sersic}}
\end{figure*}

We  check the reliability of this sub-sample by using the structural parameters derived by \cite{Tarsitano2018}. It is well known that the S\'ersic index correlates well with galaxy morphologies:  large $n_r$ is a good proxy for ETGs and vice versa for LTGs (e.g., \citealt{Fischer2019}). On the other hand, edge-on galaxies should have large ellipticity values, $\epsilon_\mathrm{r}$. In figure~\ref{fig:cum_sersic}, we show how this sub-sample populates the $\mathrm{n}_r-\epsilon_\mathrm{r}$ and $\mathrm{r}_r-\epsilon_\mathrm{r}$ planes, as well as  cumulative distribution functions (CDFs) for the S\'ersic index and the ellipticity for the two classifications schemes. 

For the ETG vs. LTG classification, we find an evident separation of each class around $\mathrm{n}_r \sim 2$ and almost no ETGs with $\epsilon_\mathrm{r} < 0.5$, as expected. Although the fraction of galaxies with high S\'ersic index classified as LTGs is $\sim$ 30\%, we note that this is due to the much larger fraction of LTGs in this sub-sample. According to the CDF, $88\%$ of the \textit{robust} ETGs have $\mathrm{n}_r > 2$, while $87\%$  of the \textit{robust} LTGs have $\mathrm{n}_r < 2$. Although the transition between ETGs and LTGs is not exactly at  $\mathrm{n}_r = 2$, the very different distributions of the ETGs and LTG samples is an indicator of the accuracy of our model predictions in real DES galaxy images. It is also interesting to note that galaxies not classified within the previous two classes, i.e., the secure intermediate candidates, show a CDF that places them in between the CDFs for the ETGs and the LTGs. 

For the face-on vs. edge-on classification scheme, we find an even sharper separation at $\epsilon_\mathrm{r} \sim 0.5$ at all radius, except for the smallest galaxies ($\mathrm{r}_r \lesssim 1.0$ arcsec), which indicates that in those cases the spatial resolution is not enough for identifying edge-on galaxies. Regarding the CDF, $87\%$ of the \textit{robust} face-on galaxies ($max (\mathrm{p^e_i}) < 0.30$) have $\epsilon_\mathrm{r} < 0.5$, while $\sim 100\%$ of the \textit{robust} edge-on ($min (\mathrm{p^e_i}) > 0.70$) have $\epsilon_\mathrm{r} > 0.5$, thus allowing us to be confident about our model predictions (figure~\ref{fig:cum_sersic}). The fact that only 0.3\% of the \textit{robust} ETGs are classified as \textit{robust} edge-on is also a good sanity check.


\subsection{VIPERS spectral classification}
\label{sec:vipers}

In this section, we compare the predictions made by our ETG vs. LTG classification  with an unsupervised machine-learning classification extracted from the VIMOS Public Extragalactic Redshift Survey (VIPERS) presented in \citet{Siudek2018}. The data release provides spectroscopic measurements and photometric properties for 86,775 galaxies. The galaxy classification is based on a Fisher Expectation-Maximization (FEM) unsupervised algorithm working in a parameter space of 12 rest-frame magnitudes and spectroscopic redshift. 

The FEM unsupervised algorithm is able to distinguish 12 classes (11 classes of  galaxies and an additional class of broad-line active galactic nuclei, AGNs). In particular, classes 1--3 host the reddest spheroidal-shape galaxies showing no sign of star formation activity and dominated by old stellar populations; classes 7--11 contain the  blue star-forming galaxies. Classes 4--6 host intermediate galaxies whose physical properties (such as colours, sSFR, stellar masses, and shapes) are intermediate between those of red, passive, and blue, active galaxies. These intermediate galaxies have more concentrated light profiles and lower gas contents than star-forming galaxies. Class 11 may consist of low-metallicity galaxies, or AGNs according to its localisation on the BPT diagram. See color-color diagrams in figure 2 of \citet{{Siudek2018}} for further details.

We  include in this comparison VIPERS galaxies  that are also present in the DES DR1 morphological catalog with an accurate spectroscopic redshift estimate and the highest membership probability to one of the classes. This subset includes 7,384 galaxies with an apparent magnitude of $\mathrm{m}_r \in [18.0, 21.5]~\mathrm{mag}$  (see figure \ref{fig:mag-distr}) and with a spectroscopic redshift distribution ranging from $0.04 < z_\mathrm{spec} < 1.46$ with a median value of $\tilde{z}_\mathrm{spec} \approx 0.55$.  Note that this is the faintest of the comparison samples.

Table \ref{tab:results} shows statistics for this sub-sample, for which  $81\%$ of the galaxies show secure classifications (i.e., $\Delta \mathrm{p} < 0.30$). Of the secure subset, $20\%$ of the galaxies are classified as \textit{robust} ETGs while 76\% are  \textit{robust} LTGs. Although the LTGs still dominate the number counts, the two classes are much more balanced than for the full DES morphological catalogue or the \cite{Tarsitano2018} sub-sample. On the other hand, the fraction of edge-on galaxies (2\%) is smaller than for the other sub-samples, and only 0.1\% of the \textit{robust} ETGs are classified as \textit{robust} edge-on.

\begin{figure}
\centering
 \includegraphics[width=\columnwidth]{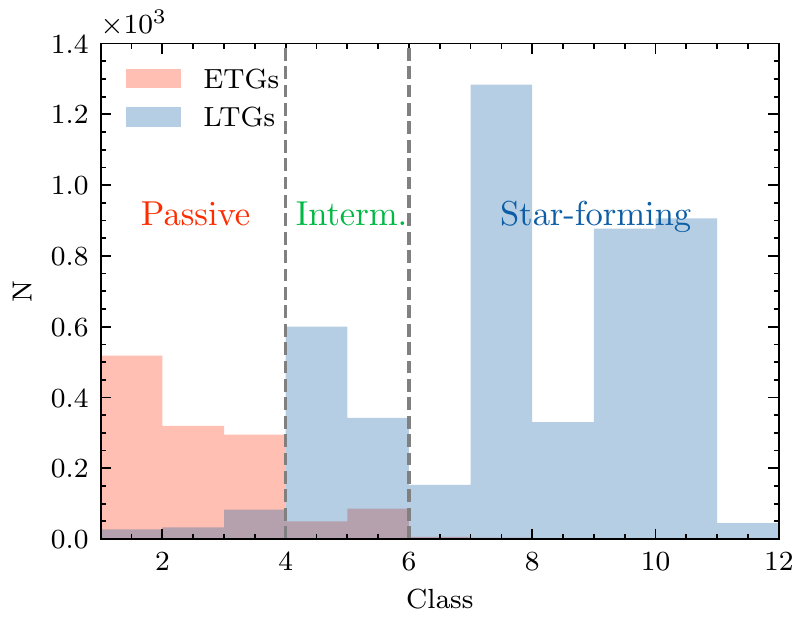}
 \caption{Number of galaxies belonging to each VIPERS class derived by \citet{Siudek2018} ranging from 1 to 12. The red histogram corresponds to the number of ETGs with $\mathrm{\tilde{p}} < 0.4$. The blue histogram shows the number of LTGs with $\mathrm{\tilde{p}} > 0.4$. The ETGs clearly dominate at classes below 4, and are negligible for classes above 6, while the LTGs dominate for classes above 6. This strong correlation demonstrates that our model is able to correctly classify original DES images, even at the fainter magnitudes.
\label{fig:vipers_classes}}
\end{figure}

In figure~\ref{fig:vipers_classes}, we show the number of galaxies belonging to each of the classes derived by \citet{Siudek2018} for the ETGs and LTGs sub-samples according to our model predictions. The ETGs clearly dominate at classes below 4, and are negligible for classes above 6. On the other hand, the LTGs dominate for classes above 6, with a very small fraction with classes 1--3. The intermediate classes are composed of a mix of ETGs and LTGs, but mainly populated by LTGs. This strong correlation nicely demonstrates that our model is able to correctly classify original DES images, even at the fainter magnitudes.

To quantify these trends, we can consider as negatives (N) the galaxies belonging to classes 1--3 and as positives (P) the galaxies falling with the classes 4--11. By doing so we find $89\%$ of TN and a $97\%$ of TP. This translates into an accuracy classification score of $\mathrm{Acc} \approx 0.95$. We have visually inspected the FN images within the VIPERS dataset (i.e., ETGs with classes 4--11) finding that for most of them there are neither clear signs of features (such as spiral arms) nor edge-on morphologies, indicating that the ETG morphological classification might be correct regardless of their spectral classification. In the case of the FP, we noticed that for a large fraction of them there is (at least) one close companion within the field of view of the cutout that might lead to an inaccurate classification. 

\section{Conclusions}

\begin{itemize}
\item We present a morphological classification according to two schemes (a) ETGs vs. LTGs, (b) edge-on vs. face-on for  $\sim$ 27 million DES galaxies to  $\mathrm{m}_r < 21.5~\mathrm{mag}$. The classifications are based on the predictions of supervised DL models using CNNs (section \ref{sec:network}). 

\item The training sample consists of bright  ($\mathrm{m}_r < 17.7~\mathrm{mag}$) DES galaxies with a previously known morphological classification (from \citealt{Dominguez2018}) as well as their artificially redshifted counterparts described in section \ref{sec:sims}.

\item Although some of the features that distinguish ETGs and LTGs almost disappear for the fainter galaxies (figure \ref{fig:high-z_examples}) the models are able to correctly classify galaxies according to the two schemes with excellent results (accuracy > 97\%) even at the fainter magnitude bins (figure \ref{fig:model-perf} and \ref{fig:model-perf_edgeon}). 

\item We train 5 different models using $k$-folding to obtain a measurement of the classification uncertainty. About 87\% of the galaxies in the final catalogue have secure labelling for the ETG vs. LTG classification (i.e., $\Delta \mathrm{p} < 0.30$). This fraction is 73\% for the edge-on classification. 

\item The classifications on real DES faint images are consistent with other available observables, such as absolute magnitude, Sersic index $n$, ellipticity $\epsilon$ or spectral classification (section \ref{sec:results}). 

\item Our work demonstrates that machines are able to recover features hidden to the human eye and so can reliably classify faint galaxy images. The methodology adopted in this work to overcome the lack of faint labelled samples can be applied to future big data surveys such as Euclid or Vera Rubin Observatory Legacy Survey of Space and Time. 

\item  The exceptional amount of data provided by DES DR1 has allowed us to construct the largest automated morphological catalogue to date (along with the companion DES morphological catalog presented in Cheng et al.) by several orders of magnitude compared to previous works (e.g., \citealt{Dominguez2018}). This classification will be a fundamental tool for our understanding of morphological transformations across cosmic time.

\item The complete DES dataset DR2, including observations for 600 million galaxies, will be made public in early 2021. The DL models presented in this work can be directly applied to DR2,  providing accurate morphological classification for a great fraction of the galaxies with very little effort. In addition, the existence of deep fields within the DES DR2 will allow us to extend this classification to even fainter magnitude limits and to carry out crucial scientific  analysis for galaxy formation and evolution.
\end{itemize}

\section*{Acknowledgements}
This work was supported in part by NSF grant AST-1816330. HDS acknowledges support from the Centro Superior de Investigaciones Científicas PIE2018-50E099. We are grateful to R. Sheth for a careful reading of the manuscript.  

Funding for the DES Projects has been provided by the U.S. Department of Energy, the U.S. National Science Foundation, the Ministry of Science and Education of Spain, 
the Science and Technology Facilities Council of the United Kingdom, the Higher Education Funding Council for England, the National Center for Supercomputing 
Applications at the University of Illinois at Urbana-Champaign, the Kavli Institute of Cosmological Physics at the University of Chicago, 
the Center for Cosmology and Astro-Particle Physics at the Ohio State University,
the Mitchell Institute for Fundamental Physics and Astronomy at Texas A\&M University, Financiadora de Estudos e Projetos, 
Funda{\c c}{\~a}o Carlos Chagas Filho de Amparo {\`a} Pesquisa do Estado do Rio de Janeiro, Conselho Nacional de Desenvolvimento Cient{\'i}fico e Tecnol{\'o}gico and 
the Minist{\'e}rio da Ci{\^e}ncia, Tecnologia e Inova{\c c}{\~a}o, the Deutsche Forschungsgemeinschaft and the Collaborating Institutions in the Dark Energy Survey. 

The Collaborating Institutions are Argonne National Laboratory, the University of California at Santa Cruz, the University of Cambridge, Centro de Investigaciones Energ{\'e}ticas, 
Medioambientales y Tecnol{\'o}gicas-Madrid, the University of Chicago, University College London, the DES-Brazil Consortium, the University of Edinburgh, 
the Eidgen{\"o}ssische Technische Hochschule (ETH) Z{\"u}rich, 
Fermi National Accelerator Laboratory, the University of Illinois at Urbana-Champaign, the Institut de Ci{\`e}ncies de l'Espai (IEEC/CSIC), 
the Institut de F{\'i}sica d'Altes Energies, Lawrence Berkeley National Laboratory, the Ludwig-Maximilians Universit{\"a}t M{\"u}nchen and the associated Excellence Cluster Universe, 
the University of Michigan, NFS's NOIRLab, the University of Nottingham, The Ohio State University, the University of Pennsylvania, the University of Portsmouth, 
SLAC National Accelerator Laboratory, Stanford University, the University of Sussex, Texas A\&M University, and the OzDES Membership Consortium.

Based in part on observations at Cerro Tololo Inter-American Observatory at NSF's NOIRLab (NOIRLab Prop. ID 2012B-0001; PI: J. Frieman), which is managed by the Association of Universities for Research in Astronomy (AURA) under a cooperative agreement with the National Science Foundation.

The DES data management system is supported by the National Science Foundation under Grant Numbers AST-1138766 and AST-1536171.
The DES participants from Spanish institutions are partially supported by MICINN under grants ESP2017-89838, PGC2018-094773, PGC2018-102021, SEV-2016-0588, SEV-2016-0597, and MDM-2015-0509, some of which include ERDF funds from the European Union. IFAE is partially funded by the CERCA program of the Generalitat de Catalunya.
Research leading to these results has received funding from the European Research
Council under the European Union's Seventh Framework Program (FP7/2007-2013) including ERC grant agreements 240672, 291329, and 306478.
We  acknowledge support from the Brazilian Instituto Nacional de Ci\^encia
e Tecnologia (INCT) do e-Universo (CNPq grant 465376/2014-2).

This manuscript has been authored by Fermi Research Alliance, LLC under Contract No. DE-AC02-07CH11359 with the U.S. Department of Energy, Office of Science, Office of High Energy Physics.

\section*{Data availability}
The DES Y1 morphological catalog is available in the Dark Energy Survey Data Management (DESDM) system at the National Center for Supercomputing Applications (NCSA) at the University of Illinois and can be accessed at \url{https://des.ncsa.illinois.edu/releases/other/morphCNN}. The pipeline used to construct the DES Y1 morphological catalog will be shared on request to the corresponding author.



\bibliographystyle{mnras}
\bibliography{refs}


\appendix
\section{Examples of galaxy images}
\label{appendix}
In the appendix we show some examples of real DES galaxies classified by our morphological catalogue in different magnitude bins. Note that our models are able to correctly classify the fainter galaxies, despite the fact that the noise significantly degrades their morphological features.  

\begin{figure*}
\centering
 \includegraphics[width=2\columnwidth]{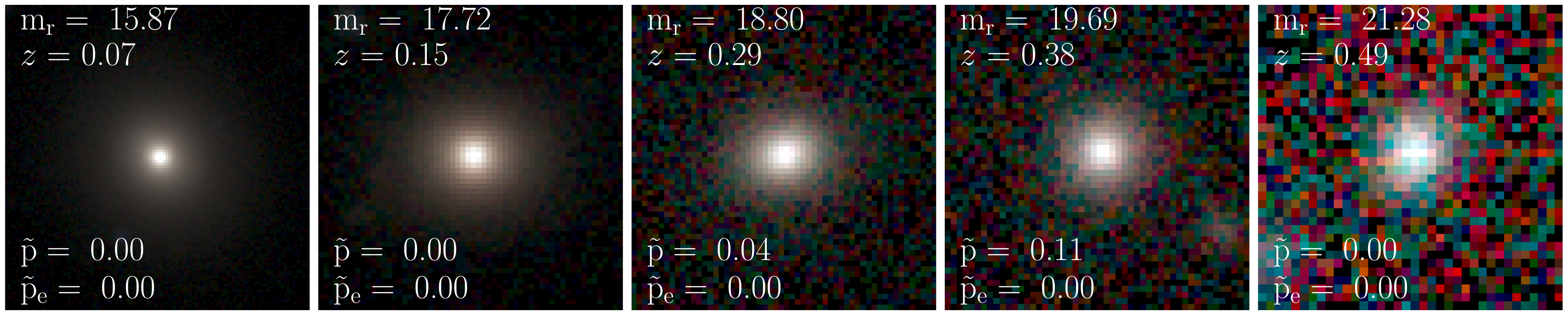}
  \includegraphics[width=2\columnwidth]{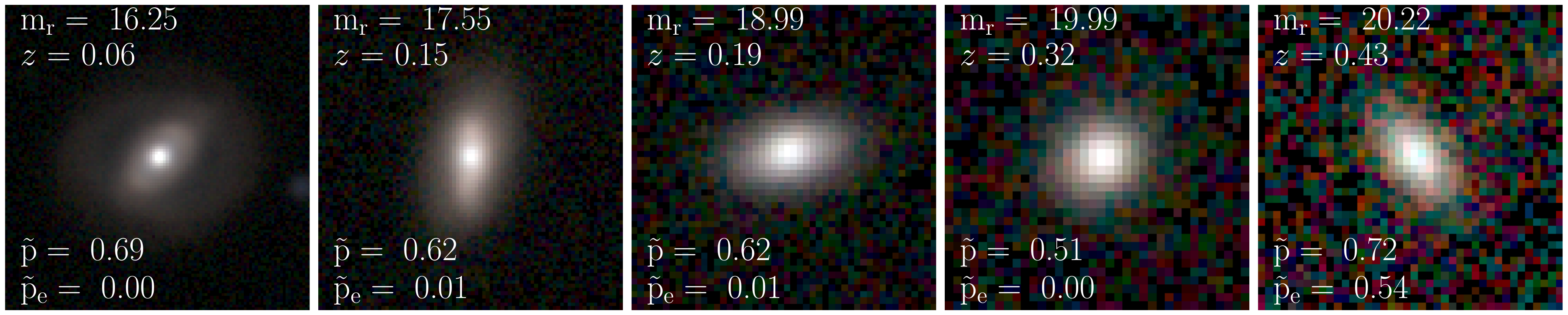}
 \includegraphics[width=2\columnwidth]{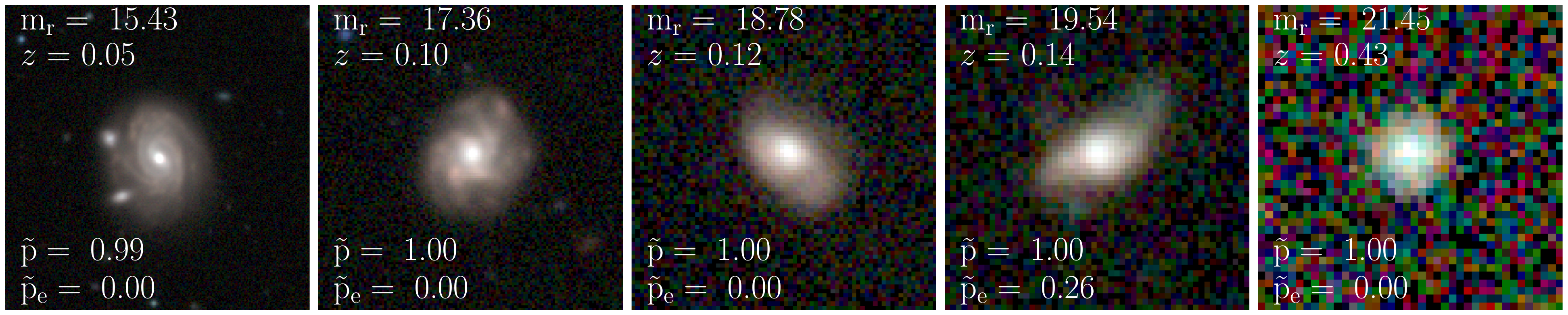}
  \includegraphics[width=2\columnwidth]{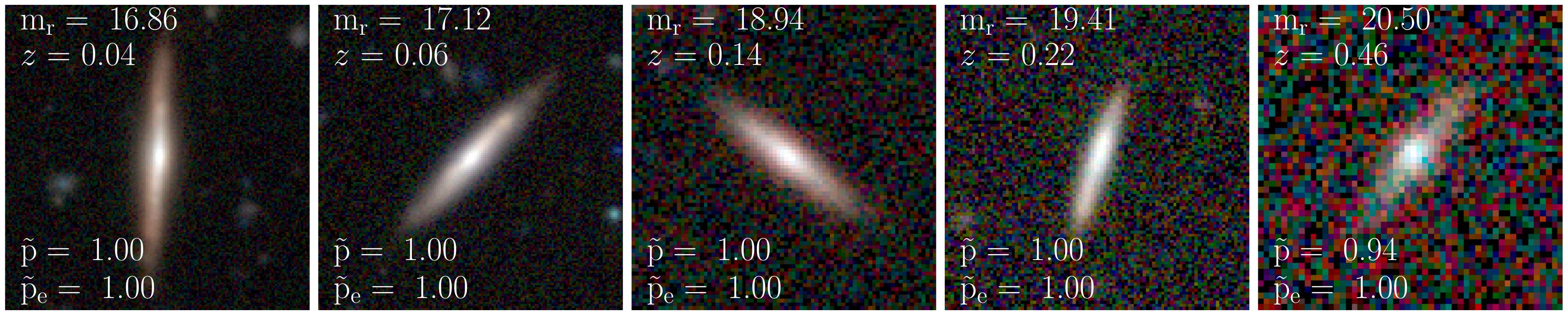}
 \includegraphics[width=2\columnwidth]{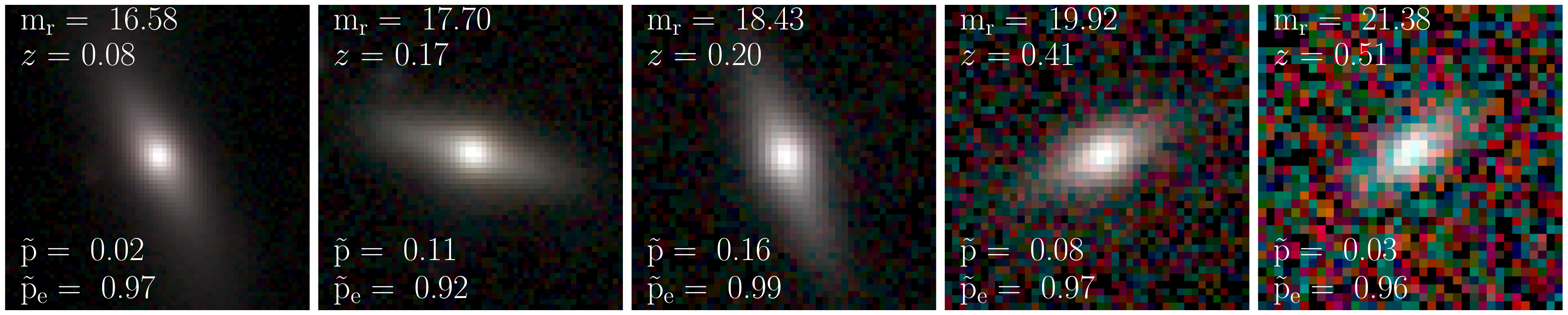}
 \caption{Examples of DES galaxies classified as \textit{robust} ETGs, secure intermediates, \textit{robust} LTGs, \textit{robust} edge-on and \textit{robust} edge-on ETGs (from top to bottom). Each column corresponds to a magnitude bin, fainter towards the right. The cutouts include the redshift and apparent magnitude of the galaxies from DES DR1 catalogue, as well as our median CNN-derived probabilities of being LTG ($\mathrm{\tilde{p}}$) and edge-on ($\mathrm{\tilde{p}_e}$).}
 \label{fig:cutouts}
\end{figure*}

\section*{Affiliations}
$^{1}$ Department of Physics and Astronomy, University of Pennsylvania, Philadelphia, PA 19104, USA\\
$^{2}$ IFCA, Instituto de F\'{\i}sica de Cantabria (UC-CSIC), Av. de Los Castros s/n, 39005 Santander, Spain\\
$^{3}$ Institute of Space Sciences (ICE, CSIC),  Campus UAB, Carrer de Can Magrans, s/n,  08193 Barcelona, Spain\\
$^{4}$ LERMA, Observatoire de Paris, PSL Research University, CNRS, Sorbonne Universit\'es, UPMC Univ. Paris 06,F-75014 Paris, France\\
$^{5}$ Univerist\'e de Paris, 5 Rue Thomas Mann - 75013, Paris, France\\
$^{6}$ Departamento de Astrof\'isica, Universidad de La Laguna, E-38206 La Laguna, Tenerife, Spain\\
$^{7}$ Instituto de Astrofisica de Canarias, E-38205 La Laguna, Tenerife, Spain\\
$^{8}$ SCIPP, University of California, Santa Cruz, CA 95064, USA\\
$^{9}$ Physics Department, 2320 Chamberlin Hall, University of Wisconsin-Madison, 1150 University Avenue Madison, WI  53706-1390\\
$^{10}$ Departamento de F\'isica Matem\'atica, Instituto de F\'isica, Universidade de S\~ao Paulo, CP 66318, S\~ao Paulo, SP, 05314-970, Brazil\\
$^{11}$ Laborat\'orio Interinstitucional de e-Astronomia - LIneA, Rua Gal. Jos\'e Cristino 77, Rio de Janeiro, RJ - 20921-400, Brazil\\
$^{12}$ Fermi National Accelerator Laboratory, P. O. Box 500, Batavia, IL 60510, USA\\
$^{13}$ Instituto de Fisica Teorica UAM/CSIC, Universidad Autonoma de Madrid, 28049 Madrid, Spain\\
$^{14}$ Institute of Cosmology and Gravitation, University of Portsmouth, Portsmouth, PO1 3FX, UK\\
$^{15}$ CNRS, UMR 7095, Institut d'Astrophysique de Paris, F-75014, Paris, France\\
$^{16}$ Sorbonne Universit\'es, UPMC Univ Paris 06, UMR 7095, Institut d'Astrophysique de Paris, F-75014, Paris, France\\
$^{17}$ Department of Physics \& Astronomy, University College London, Gower Street, London, WC1E 6BT, UK\\
$^{18}$ Universidad de La Laguna, Dpto. AstrofÃ­sica, E-38206 La Laguna, Tenerife, Spain\\
$^{19}$ Department of Astronomy, University of Illinois at Urbana-Champaign, 1002 W. Green Street, Urbana, IL 61801, USA\\
$^{20}$ National Center for Supercomputing Applications, 1205 West Clark St., Urbana, IL 61801, USA\\
$^{21}$ Institut de F\'{\i}sica d'Altes Energies (IFAE), The Barcelona Institute of Science and Technology, Campus UAB, 08193 Bellaterra (Barcelona) Spain\\
$^{22}$ Center for Cosmology and Astro-Particle Physics, The Ohio State University, Columbus, OH 43210, USA\\
$^{23}$ Jodrell Bank Center for Astrophysics, School of Physics and Astronomy, University of Manchester, Oxford Road, Manchester, M13 9PL, UK\\
$^{24}$ University of Nottingham, School of Physics and Astronomy, Nottingham NG7 2RD, UK\\
$^{25}$ INAF-Osservatorio Astronomico di Trieste, via G. B. Tiepolo 11, I-34143 Trieste, Italy\\
$^{26}$ Institute for Fundamental Physics of the Universe, Via Beirut 2, 34014 Trieste, Italy\\
$^{27}$ Observat\'orio Nacional, Rua Gal. Jos\'e Cristino 77, Rio de Janeiro, RJ - 20921-400, Brazil\\
$^{28}$ Department of Physics, University of Michigan, Ann Arbor, MI 48109, USA\\
$^{29}$ Centro de Investigaciones Energ\'eticas, Medioambientales y Tecnol\'ogicas (CIEMAT), Madrid, Spain\\
$^{30}$ Department of Physics, IIT Hyderabad, Kandi, Telangana 502285, India\\
$^{31}$ Institute of Theoretical Astrophysics, University of Oslo. P.O. Box 1029 Blindern, NO-0315 Oslo, Norway\\
$^{32}$ Institut d'Estudis Espacials de Catalunya (IEEC), 08034 Barcelona, Spain\\
$^{33}$ Kavli Institute for Cosmological Physics, University of Chicago, Chicago, IL 60637, USA\\
$^{34}$ Department of Physics, Stanford University, 382 Via Pueblo Mall, Stanford, CA 94305, USA\\
$^{35}$ Kavli Institute for Particle Astrophysics \& Cosmology, P. O. Box 2450, Stanford University, Stanford, CA 94305, USA\\
$^{36}$ SLAC National Accelerator Laboratory, Menlo Park, CA 94025, USA\\
$^{37}$ D\'{e}partement de Physique Th\'{e}orique and Center for Astroparticle Physics, Universit\'{e} de Gen\`{e}ve, 24 quai Ernest Ansermet, CH-1211 Geneva, Switzerland\\
$^{38}$ Department of Physics, ETH Zurich, Wolfgang-Pauli-Strasse 16, CH-8093 Zurich, Switzerland\\
$^{39}$ School of Mathematics and Physics, University of Queensland,  Brisbane, QLD 4072, Australia\\
$^{40}$ Santa Cruz Institute for Particle Physics, Santa Cruz, CA 95064, USA\\
$^{41}$ Department of Physics, The Ohio State University, Columbus, OH 43210, USA\\
$^{42}$ Faculty of Physics, Ludwig-Maximilians-Universit\"at, Scheinerstr. 1, 81679 Munich, Germany\\
$^{43}$ Max Planck Institute for Extraterrestrial Physics, Giessenbachstrasse, 85748 Garching, Germany\\
$^{44}$ Universit\"ats-Sternwarte, Fakult\"at f\"ur Physik, Ludwig-Maximilians Universit\"at M\"unchen, Scheinerstr. 1, 81679 M\"unchen, Germany\\
$^{45}$ Lawrence Berkeley National Laboratory, Berkeley, CA 94720, USA\\
$^{46}$ Australian Astronomical Optics, Macquarie University, North Ryde, NSW 2113, Australia\\
$^{47}$ Lowell Observatory, 1400 Mars Hill Rd, Flagstaff, AZ 86001, USA\\
$^{48}$ Instituci\'o Catalana de Recerca i Estudis Avan\c{c}ats, E-08010 Barcelona, Spain\\
$^{49}$ Institute of Astronomy, University of Cambridge, Madingley Road, Cambridge CB3 0HA, UK\\
$^{50}$ Department of Astrophysical Sciences, Princeton University, Peyton Hall, Princeton, NJ 08544, USA\\
$^{51}$ Department of Physics and Astronomy, Pevensey Building, University of Sussex, Brighton, BN1 9QH, UK\\
$^{52}$ School of Physics and Astronomy, University of Southampton,  Southampton, SO17 1BJ, UK\\
$^{53}$ Computer Science and Mathematics Division, Oak Ridge National Laboratory, Oak Ridge, TN 37831\\

\bsp	
\label{lastpage}
\end{document}